\documentclass[reprint,aip,jmp,showpacs,onecolumn,nofootinbib]{revtex4-1}

\usepackage{amsthm,amssymb,amsmath}				
\usepackage{enumerate}							
\usepackage{graphicx}							
\usepackage[colorlinks = true,linkcolor = blue, 
            urlcolor  = blue, citecolor = blue, 
            anchorcolor = blue]{hyperref}       

\begin{document}

\title{
Supersymmetric quantum mechanics and coherent states for 
a deformed oscillator with position-dependent effective mass
}

\author{Bruno G.\ da Costa}
\email{bruno.costa@ifsertao-pe.edu.br}
\author{Genilson A. C. da Silva}
\email{cardoso.genilson@outlook.com}
\affiliation{Instituto Federal de Educa\c{c}\~ao, Ci\^encia e Tecnologia do Sert\~ao Pernambucano,
             Rua Maria Luiza de Ara\'ujo Gomes Cabral s/n, 56316-686 Petrolina, Pernambuco, Brazil}
\author{Ignacio S.\ Gomez}
\email{ignacio.sebastian@ufba.br}
\affiliation{Instituto de Fisica, Universidade Federal da Bahia,
             R.\ Barao de Jeremoabo s/n, 40170-115 Salvador, Bahia, Brazil}
\date{\today}

\begin{abstract}
We study the classical and quantum oscillator in the 
context of a non-additive (deformed) displacement operator, 
associated with a position-dependent effective mass, 
by means of the supersymmetric formalism.
From the supersymmetric partner Hamiltonians 
and the shape invariance technique
we obtain the eigenstates and the eigenvalues along with 
the ladders operators, thus showing a preservation of the 
supersymmetric structure in terms of the deformed counterpartners.
The deformed space in supersymmetry allows to characterize
position-dependent effective mass, uniform field interactions and  
to obtain a generalized uncertainty relation (GUP)
that behaves as a distinguishability measure 
for the coherent states, these latter satisfying
a periodic evolution of the GUP corrections.
\end{abstract}


\maketitle 

\section{Introduction}

The factorization method is an operational procedure that 
enables us to answer questions about eigenvalue problems which 
are of importance to physicists.\cite{Dong_2007} 
This algebraic method was first introduced by Schr\"odinger 
in the problem of the quantum harmonic oscillator,\cite{Schrodinger-1940}
in which the eigenstates are obtained
from the application of the creation 
and annihilation operators to the ground state.
The introduction of the supersymmetry to study quantum systems 
can be understood as a generalization of this idea.
More specifically, Witten\cite{Witten-1981,Cooper-Freedman-1983} studied
properties of symmetry within the context of string theory, 
in order to unify fermionic and bosonic systems, in what became known 
as supersymmetric quantum mechanics (supersymmetry), 
commonly called SUSY. 
The algebra involved in SUSY is a Lie algebra with a combination 
of commutation and anti-commutation relationships.
SUSY has been applied in several contexts of quantum mechanics: 
infinite square potential,\cite{Cooper-Khare-Sukhatme-1995}
hydrogen atom,\cite{RosasOrtiz-1998}
Morse,\cite{Benedict-Molnar-1999}
Lennard-Jones,\cite{Borges-Filho-2001}
Rosen-Morse\cite{Compean-Kirchbach-2005}
and P\"oschl-Teller\cite{Dong_Lemus_2002}
potentials, among others.

It is also attributed to Schr\"odinger for introducing 
the idea of coherent states in a seminal paper in 1926 
about the simple harmonic oscillator.\cite{Schrodinger-1926}
In that work, a superposition of quantum states is built 
to reproduce the dynamics of the corresponding classical analog.  
Glauber,\cite{Glauber-1963} Klauder\cite{Klauder-1963a,Klauder-1963b} 
and Sudarshan\cite{Sudarshan-1963} in the early 1960s 
were the pioneers to apply the idea of coherent states 
in quantum optics.
The term `coherent states' was first used by
Glauber in his work on electromagnetic radiation, 
and defined as the eigenstates of the annihilation operator 
for the quantum harmonic oscillator.
In addition, it is also shown the coherent states for 
the harmonic oscillator satisfy the minimization 
of the Heisenberg's uncertainty principle.\cite{Gazeau-2009}

Complementary, another important issue in quantum mechanics is 
the concept of position-dependent mass (PDM), which has attracted 
the attention along the decades due its wide applicability in: 
semiconductor heterostructures,
\cite{Bastard-1975,
vonroos_1983,
BenDaniel-Duke-1966,
Gora-Williams-1969,
Zhu-Kroemer-1983,
Li-Kuhn-1993,
Morrow-Brownstein-1984,
Mustafa-Mazharimousavi-2007}
nonlinear optics,\cite{Khordad}
quantum liquids,\cite{Saavedra_1994} 
many-body theory,\cite{Bencheikh-2004} 
molecular physics,\cite{Christiansen-Cunha-2014}
Wigner functions,\cite{Cherroud-2017} 
quantum information,\cite{Yanez-Navarro}
relativistic quantum mechanics,\cite{Glasser-2020} 
Dirac equation,\cite{Jia-SouzaDutra-2008}
superintegrable systems,\cite{Ranada-2016} 
nuclear physics,\cite{Alimohammadi-Hassanabadi-Zare-2017}
magnetic monopoles,\cite{Jesus-2019}
Landau quantization,\cite{Algadhi-Mustafa-2020}
factorization and supersymmetry 
methods,\cite{Plastino-etal-1999,
              Amir-Iqbal-2016,
              Karthiga-2018,
              Bravo-PRD-2016,
              Mustafa-2020}
coherent states,\cite{Ruby-Senthilvelan-2010,
Amir-Iqbal-2015,Amir-Iqbal-2016-CS,Tchoffo-2019}
etc.

Furthermore, effects of the gravitational field in quantum mechanics 
have been characterized by generalizations of the standard commutation relationship 
between the position and the linear momentum, 
giving place to the generalized 
uncertainty principles (GUP).
\cite{Kem-1994,
      Benczik-1994,
      Pedram-2012,
      Hossenfelder-2013,
      Bosso-2018,
      Costa-Filho-2016,
      daCosta-Gomez-Portesi-2020,
      Merad-2020}
In this context, some theoretical frameworks 
have been developed to mimic
the effect of the PDM by means of 
deformed algebraic structures.
\cite{Costa-Filho-2016,
daCosta-Gomez-Portesi-2020,Merad-2020}
One of these formulations 
is derived from a translation operator 
that causes non-additive displacements of the type
$\hat{\mathcal{T}}_\gamma (\varepsilon)|x\rangle 
= | x + (1 + \gamma x) \varepsilon \rangle$,
being $\gamma$ a deformation parameter 
with inverse length dimension.
\cite{CostaFilho-Almeida-Farias-AndradeJr-2011,
Mazharimousavi-2012,
Costa-Borges-2014,
Barbagiovanni-Costafilho-2013,
Barbagiovanni-2014,
Costa-Gomez-Santos-2020,
CostaFilho-Alencar-Skagerstam-AndradeJr-2013,
Costa-Borges-2018,
Costa-Gomez-Borges-2020,
Tchoffo-Vubangsi-Fai-2014,
Merad-etal_2019,
Arda-Server,
Aguiar-Cunha-daCosta-CostaFilho-2020,
CostaFilho-Oliveira-Aguiar-DaCosta-2021}
This translation operator leads to a position-dependent 
linear momentum operator $\hat{p}_\gamma$ that generates
non-additive translations. 
Consequently, the particle mass is 
a function of the position controlled 
by the parameter $\gamma$.
An associated deformed position operator $\hat{x}_\gamma$ 
maps the Hamiltonian of a particle with a PDM 
into another Hamiltonian with constant mass.
In the displacement-operator formalism, 
the time-independent Schr\"odinger equation 
can be expressed using a deformed derivative operator 
$D_\gamma = (1 + \gamma x) \textrm{d}/\textrm{d}x$, 
which results physically equivalent
to introduce a particle with a PDM. 
Typical problems of quantum mechanics have been solved 
within this approach: 
infinite and finite 
square potential wells,\cite{CostaFilho-Almeida-Farias-AndradeJr-2011,
      Mazharimousavi-2012,Costa-Borges-2014} 
quantum dots and wells,
\cite{Barbagiovanni-Costafilho-2013,Barbagiovanni-2014}
quasi-periodic\cite{Costa-Gomez-Santos-2020}
and Coulomb-like potentials,\cite{Arda-Server}
harmonic oscillator,\cite{CostaFilho-Alencar-Skagerstam-AndradeJr-2013,
	  Costa-Borges-2018,
	  Costa-Gomez-Borges-2020,
	  Tchoffo-Vubangsi-Fai-2014,
      Merad-etal_2019}
Dirac fermions in graphene\cite{Aguiar-Cunha-daCosta-CostaFilho-2020}
and two dimensional electron gas.\cite{CostaFilho-Oliveira-Aguiar-DaCosta-2021}
It can be shown that the energy spectrum of the 
deformed harmonic oscillator corresponds
to the Morse oscillator, i.e., an anharmonic oscillator.

In spite of the several applications of 
the position-dependent translation operator formalism,
the SUSY method has not been used yet to characterize 
the deformed oscillator. 
The goal of this work is to fulfill this gap
by extending the SUSY method  
to the quantum and classical harmonic oscillator 
with PDM within the formalism 
of non-additive operators.
\cite{CostaFilho-Alencar-Skagerstam-AndradeJr-2013,Costa-Borges-2018}.
We also calculate the corresponding coherent states, 
that reproduce the trajectory in the phase space of 
their respective non-additive classic analogues, investigated recently in 
Ref.~\onlinecite{Costa-Borges-2018}.

The paper is organized as follows.
In Section \ref{sec:classical-and-quantum-deformed-formalism}, 
we review the classical and quantum mechanics
of the displacement operator approach.
Section \ref{sec:factorization-method}
is devoted to the study of the factorization method and SUSY 
for the classical and quantum deformed oscillator,
previously introduced in 
Refs.~\onlinecite{CostaFilho-Alencar-Skagerstam-AndradeJr-2013,
	  Costa-Borges-2018}.
Then, in Section \ref{sec:coherent-states} we calculate the
coherent states in the position representation. 
For these quasi-classical states, we investigate 
the time-evolution of the position and the linear momentum 
along with the uncertainty relation.
Finally, in Section \ref{sec:final-remarks}
we draw the conclusions and some perspectives are outlined.

\section{\label{sec:classical-and-quantum-deformed-formalism}
		 Deformed classical and quantum mechanics for 
  		 position-dependent mass}

\subsection{Deformed classical formalism}

Let us initially address the problem of a
harmonic oscillator with PDM in the
classical formalism, whose Hamiltonian is 
\begin{equation}
\label{eq:hamiltonian_x_p}
 \mathcal{H}(x , p) = \frac{p^2}{2m(x)} +  V(x).
\end{equation}
The equation of motion is
\begin{equation}
\label{eq:motion-equation-m(x)}
	m(x)\ddot{x}+\frac{1}{2}m'(x)\dot{x}^2 = F(x)
\end{equation}
with $F(x) = -\textrm{d}V/\textrm{d}x$ 
the conservative force acting on the particle,
$\dot{x} = \textrm{d}x/\textrm{d}t$ the velocity, 
$\ddot{x} = \textrm{d}^2 x/\textrm{d}t^2$ the acceleration,
and $m'(x) = \textrm{d}m/\textrm{d}x$ the mass gradient.

In particular, for the mass function
\begin{equation}
\label{eq:m(x)}
 m(x) = \frac{m_0}{(1+\gamma x)^2},
 \qquad (x > -1/\gamma )
\end{equation}
the equation of motion (\ref{eq:motion-equation-m(x)})
may be conveniently rewritten as
\begin{equation}
\label{eq:second_newton_law_generalized}
       m_0 \mathcal{D}^2_{\gamma} x (t) = F(x),
\end{equation}
i.e., a deformed Newton's law for a space with 
nonlinear displacements governed by 
the (nonlinear) deformed derivative operator
$
\mathcal{D}_{\gamma} x(t) = 
\frac{1}{1+\gamma x} \frac{\textrm{d}x}{\textrm{d}t}.
$
\cite{Costa-Borges-2014,Costa-Gomez-Borges-2020}

The point canonical transformation\cite{Costa-Borges-2018}
\begin{subequations}
\label{eq:classical-dynamical-variables}
\begin{equation}
x_\gamma = \int^x \sqrt{\frac{m(y)}{m_0}}\textrm{d}y
		 = \frac{\ln (1 + \gamma x)}{\gamma}\\
\end{equation}
and
\begin{equation}
\Pi_\gamma = \sqrt{\frac{m_0}{m(x)}} p 
    	   = (1+\gamma x)p
\end{equation}
\end{subequations}
maps the Hamiltonian (\ref{eq:hamiltonian_x_p}) 
of a particle with PDM $m(x)$ (\ref{eq:m(x)}) 
in the usual phase space $(x,p)$ into another
Hamiltonian of a particle with a constant mass
$m_0$ represented in the deformed phase space
$(x_\gamma, \Pi_\gamma)$,
\begin{equation}
\label{eq:hamiltonian_x_gamma_p_gamma}
\mathcal{K}(x_\gamma, \Pi_\gamma) = 
\frac{1}{2m_0}\Pi_\gamma^2 + U({x_\gamma}),
\end{equation}
with $U(x_\gamma) = V(x(x_\gamma))$ 
the potential in the deformed space-coordinate $x_\gamma$.
The generalized displacement of a PDM $m(x)$
in a usual space ($\textrm{d}_\gamma x$) 
is mapped into a constant mass $m_0$
in a deformed space with usual displacement 
($\textrm{d} x_\gamma$):
$\displaystyle \textrm{d}_\gamma x
    \equiv \frac{\textrm{d}x}{1+\gamma x} = \textrm{d} x_\gamma$. 
Both representations (\ref{eq:hamiltonian_x_p})
and (\ref{eq:hamiltonian_x_gamma_p_gamma}) 
coincide for $m(x)=m_0$ ($\gamma =0$).

\subsection{Deformed quantum formalism}

In the quantization of systems with PDM, the mass function $m(\hat{x})$ 
and the linear momentum $\hat{p}$ are not commuting operators,
which leads to the problem of ordering ambiguity in the definition 
of the kinetic energy operator.
A general form for a Hermitian kinetic energy operator
of a particle with variable mass
in one-dimensional was introduced by von Roos
\cite{vonroos_1983} 
\begin{equation}
\label{eq:general-kinetic-operator-pdm}
\hat{T}(\hat{x}, \hat{p}) =
		\frac{1}{4} \left\{
			[m(\hat{x})]^{-\xi}\hat{p}
			[m(\hat{x})]^{-1+\xi+\zeta}
			\hat{p}[m(\hat{x})]^{-\zeta} 
		+[m(\hat{x})]^{-\zeta}\hat{p}
		[m(\hat{x})]^{-1+\xi+\zeta}
		\hat{p}[m(\hat{x})]^{-\xi} 
		\right\},
\end{equation}
with $\xi$ and $\zeta$ named ambiguity parameters.

Several proposals for the kinetic energy operator
are particular case of (\ref{eq:general-kinetic-operator-pdm}).
We point out some them:
Ben Daniel and Duke,\cite{BenDaniel-Duke-1966} ($\xi = \zeta = 0$)
Gora and Williams,\cite{Gora-Williams-1969} ($\xi = 1$, $\zeta = 0$)
Zhu and Kroemer,\cite{Zhu-Kroemer-1983} ($\xi = \zeta = \frac{1}{2}$)
Li and Kuhn.\cite{Li-Kuhn-1993} ($\xi = 0, \zeta = \frac{1}{2}$)
In according to Morrow and Brownstein,\cite{Morrow-Brownstein-1984}
the case $\xi = \zeta$ satisfies the conditions of continuity 
of the wave function at the boundaries of 
a heterojunction in crystals.
In particular, Mustafa and Mazharimousavi 
\cite{Mustafa-Mazharimousavi-2007}
have shown that the case $\xi  = \zeta = \frac{1}{4}$  
allows the mapping of a quantum Hamiltonian with PDM
into a Hamiltonian with constant mass
by means a point canonical transformation
that is independent of the potential $V(\hat{x})$ 
of the particle.
Considering the quantum Hamiltonian
\begin{equation}   
\label{eq:hamiltonian-MM}
 \hat{H}(\hat{x}, \hat{p}) = \frac{1}{2} 
	[m(\hat{x})]^{-\frac{1}{4}} 
	\hat{p} [m(\hat{x})]^{-\frac{1}{2}} \hat{p} 
	[m(\hat{x})]^{-\frac{1}{4}} 
 	+ V(\hat{x}),
\end{equation}   
the time-independent Schr\"odinger equation 
$\hat{H}| \psi \rangle = E| \psi \rangle $
in the representation $\{ |\hat{x}\rangle \}$ becomes
\begin{equation}
\label{eq:SE-MM-m(x)}
 \left[ -\frac{\hbar^2}{2m_0} \sqrt[4]{\frac{m_0}{m(x)}}
 \frac{{\textrm{d}}}{{\textrm{d}}x} \sqrt{\frac{m_0}{m(x)}} 
 \frac{{\textrm{d}}}{{\textrm{d}}x}\sqrt[4]{\frac{m_0}{m(x)}} 
 + V(x) \right] \psi (x)
 = E\psi(x),
\end{equation}
where $\psi (x)$ is the wavefunction solution and
$m(x) = m_0$ recovers the standard Schr\"odinger equation.

The solution of Eq.~(\ref{eq:SE-MM-m(x)}) can be performed 
through different approaches.
For instance, from the transformation
$\psi (x) = \sqrt[4]{m_0/m(x)} \varphi (x)$	
and the mass function (\ref{eq:m(x)}),
Eq.~(\ref{eq:SE-MM-m(x)}) may be conveniently rewritten
as a time-independent 
{\it deformed Schr\"odinger equation}
\cite{CostaFilho-Almeida-Farias-AndradeJr-2011,Costa-Borges-2018}
\begin{equation}
\label{eq:deformed-SE}
	-\frac{\hbar^2}{2m_0} D_{\gamma}^2 \varphi (x)
	+V(x) \varphi (x) = E \varphi (x),
\end{equation}
where
$D_{\gamma} =(1+\gamma x) \frac{{\textrm{d}}}{{\textrm{d}}x}$
is a (linear) deformed derivative operator.
\cite{CostaFilho-Almeida-Farias-AndradeJr-2011,Costa-Gomez-Borges-2020}
The Eq.~(\ref{eq:deformed-SE}) corresponds to
a Schr\"odinger-like equation for $\varphi (x)$ 
expressed in terms of the non-Hermitian Hamiltonian operator
\begin{equation}
	\hat{H}_\gamma = \frac{1}{2m_0}\hat{p}_\gamma^2 + V(\hat{x})
\end{equation}
and 
$\hat{p}_\gamma \equiv (1+\gamma \hat{x})\hat{p}
= -\textrm{i}\hbar D_{\gamma}$  
a deformed non-Hermitian momentum operator,
which satisfies the commutator relation
$[\hat{x}, \hat{p}_\gamma] = 
\textrm{i}\hbar(\hat{1} + \gamma \hat{x}).
$\cite{CostaFilho-Almeida-Farias-AndradeJr-2011,Costa-Borges-2018}
The eigenfunctions $\varphi(x)$ are normalized 
by means of a deformed inner product
$\langle \varphi_1 | \varphi_2 \rangle 
= \int_{x_i}^{x_f} \varphi_1^{\ast} (x)
\varphi_2 (x){\textrm{d}}_\gamma x$,
so that the probability density for the eigenstates is
$
\rho (x) = \psi^{\ast}(x) \psi(x) = 
\frac{1}{1+\gamma x} {\varphi}^{\ast}(x){\varphi}(x).
$

Using the variable change 
$x\rightarrow x_\gamma =\gamma^{-1} \ln(1+\gamma x)$,
Eq.~(\ref{eq:deformed-SE}) can be rewritten in a deformed space
$x_\gamma$ as
\begin{equation}
\label{eq:SE-deformed-space}
	-\frac{\hbar^2}{2m_0} \frac{\textrm{d}^2 
	\phi (x_{\gamma})}{\textrm{d} x_\gamma^2}
	+U(x_{\gamma}) \phi (x_{\gamma}) = E\phi (x_{\gamma}),
\end{equation}
where $\phi (x_\gamma) = \varphi (x(x_\gamma))$ and
$U(x_\gamma) = V(x(x_\gamma))$.
Therefore, the wave equation for the field $\psi (x)$ of 
a system with PDM (\ref{eq:m(x)}) in the standard space $\{|\hat{x} \rangle\}$
is mapped into another wave equation for the field $\phi (x_\gamma)$ 
in a deformed space $\{|\hat{x}_\gamma \rangle\}$.
Of course, the quantum Hamiltonian associated with the Schr\"odinger
wave equation (\ref{eq:SE-deformed-space}) is
\begin{equation}
\hat{K}(\hat{x}_\gamma,\hat{\Pi}_\gamma) 
 = \frac{1}{2m_0}\hat{\Pi}_\gamma^2 + U(\hat{x}_\gamma),
\end{equation}
and it can be obtained applying the following point canonical transformation
$(\hat{x}, \hat{p}) \rightarrow (\hat{x}_\gamma, \hat{\Pi}_\gamma)$
on the quantum Hamiltonian (\ref{eq:hamiltonian-MM}).
The {\it space and linear pseudo-momentum operators}
are given respectively by 
\cite{Mustafa-Mazharimousavi-2007,Costa-Borges-2014}
\begin{subequations}
\begin{align}
\label{eq:hat-gamma-Pi_gamma}
\hat{x}_\gamma &=
\frac{\ln (\hat{1} + \gamma \hat{x})}{\gamma} 
\\
\hat{\Pi}_\gamma &
	=\sqrt[4]{\frac{m_0}{m(\hat{x})}}\,\hat{p}\,\sqrt[4]{\frac{m_0}{m(\hat{x})}}
	\nonumber \\
	&=(\hat{1} + \gamma \hat{x})^{1/2} \hat{p}
	(\hat{1} + \gamma \hat{x})^{1/2}
 	\nonumber \\
	&=\frac{(\hat{1} + \gamma \hat{x}) \hat{p}}{2} 
      +\frac{\hat{p} (\hat{1} + \gamma \hat{x})}{2},
\end{align}
\end{subequations}
with $[\hat{x}_\gamma, \hat{\Pi}_\gamma] = i\hbar \hat{1}$,
such that, $\hat{x}_\gamma$ and $\hat{\Pi}_\gamma$
are Hermitian operators and canonically conjugated.

The deformed momentum operator $\hat{\Pi}_\gamma$ 
also allows to express the Hamiltonian operator 
(\ref{eq:hamiltonian-MM}) for the mass function (\ref{eq:m(x)})
in the simplified form 
$\hat{H} (\hat{x}, \hat{p}) 
= \frac{1}{2m_0}\hat{\Pi}_\gamma^2 (\hat{x}, \hat{p})
+ V(\hat{x}).
$
The connection between pseudo-momentum $\hat{\Pi}_\gamma$ and
non-Hermitian momentum $\hat{p}_\gamma$ is 
$
\hat{\Pi}_\gamma 
 	=\frac{1}{2} (\hat{p}^{\dagger}_\gamma + \hat{p}_\gamma).
$
Likewise, we have
$
[\hat{x}, \hat{\Pi}_\gamma] = i\hbar (1 + \gamma \hat{x}).
$
In accordance to the generalized uncertainty principle (GUP)
$(\Delta x)^2 (\Delta \Pi_\gamma)^2 \geq 
\frac{1}{4} |\langle [\hat{x}, \hat{\Pi}_\gamma] \rangle|^2,$
and so, 
\begin{equation}
\label{eq:uncertainty-relation}
\Delta x \Delta \Pi_\gamma 
\geq \frac{\hbar}{2} (1 + \gamma \langle \hat{x} \rangle).
\end{equation}
GUP has been applied in problems of quantum mechanics 
and quantum gravity, whose modified relation commutation 
between position and linear momentum depends on
one or both operators
\cite{Kem-1994,
      Benczik-1994,
      Pedram-2012,
      Hossenfelder-2013,
      Bosso-2018,
      Costa-Filho-2016,
      daCosta-Gomez-Portesi-2020,
      Merad-2020}.
Consequently, this could be interpreted as 
an effective mass dependent on the position or the linear momentum
(see references 
\onlinecite{Costa-Filho-2016,
      daCosta-Gomez-Portesi-2020,
      Merad-2020}
for more details).
Hereinafter, we focus on the case where 
the commutator between the position 
and the pseudo-momentum operators 
is a linear function on the position,
which emerges from the displacement operator method.
\cite{CostaFilho-Almeida-Farias-AndradeJr-2011}

\section{\label{sec:factorization-method}
		 Factorization method for deformed classical 
		 and quantum oscillator}

From a pedagogical point of view, before applying SUSY techniques 
to the deformed quantum oscillator we first obtain the solution 
of the classic analog by means of the factorization method.

\subsection{Factorization method for deformed classical oscillator}

The classical Hamiltonian (\ref{eq:hamiltonian_x_p})
for the quadratic potential $V(x) = \frac{1}{2}m_0\omega_0^2 x^2$
and the mass function (\ref{eq:m(x)}) is
\begin{equation}
\label{eq:hamiltonian_oscillator}
\mathcal{H}(x , p) 
	= \frac{(1+\gamma x)^2p^2}{2m_0} 
	+  \frac{1}{2}m_0\omega_0^2 x^2.
\end{equation}
The motion equation can be obtained from 
the differential equation
$\mathcal{D}_\gamma^2 x(t) = -\omega_0^2 x.$
\cite{Costa-Borges-2018}
However, here we use the factorization method 
like an alternative way.
For this purpose, we consider 
the following dynamical variable and 
its complex conjugate
\begin{subequations}
\begin{align}
\label{eq:alpha_gamma}
& \alpha_\gamma (x,p) = \sqrt{\frac{m_0 \omega_0}{2\hbar}}
				\left[ x + \frac{i}{m_0 \omega_0} (1+\gamma x) p \right],
\\
& \alpha_\gamma^{\ast} (x,p) = \sqrt{\frac{m_0 \omega_0}{2\hbar}}
				\left[ x - \frac{i}{m_0 \omega_0} (1+\gamma x) p \right].
\end{align}
\end{subequations}
Of course, we have that the position and the linear pseudo-momentum 
are respectively
$x = \sqrt{2} \sigma_0 \textrm{Re} (\alpha_\gamma)$
and
$\Pi_\gamma = \sqrt{2} (\hbar/\sigma_0) \textrm{Im} (\alpha_\gamma),$
with 
$\sigma_0 = \sqrt{\frac{\hbar}{m_0 \omega_0}}$ 
a characteristic length,
i.e., the complex number $\alpha_\gamma$
characterizes the state of the deformed harmonic oscillator.

The classical Hamiltonian (\ref{eq:hamiltonian_oscillator})
is factorized into
\begin{equation}
\mathcal{H} = 
\hbar \omega_0 \alpha_{\gamma}^{\ast} \alpha_{\gamma},
\end{equation}
with
$\alpha_\gamma$, $\alpha_\gamma^{\ast}$ and
$\mathcal{H}$ 
satisfying the Poisson brackets
\begin{subequations}
\begin{align}
& \{ \alpha_\gamma, \alpha_\gamma^{\ast} \} = \frac{1}{i\hbar}(1+\gamma x), 
\\
& \{ \alpha_\gamma, \mathcal{H} \} =  -i\omega_0 (1+\gamma x)\alpha_\gamma,  \\
& \{ \alpha_\gamma^{\ast}, \mathcal{H} \} =  i\omega_0 (1+\gamma x)\alpha_\gamma^{\ast},
\end{align}
\end{subequations}
as well as the  Jacobi identify 
\begin{equation}
\label{eq:classical-Jacobi}
\{ \{ \alpha_\gamma, \alpha_\gamma^{\ast} \}, \mathcal{H} \} +
\{ \{ \mathcal{H}, \alpha_\gamma \}, \alpha_\gamma^{\ast} \} +
\{ \{ \alpha_\gamma^{\ast}, \mathcal{H} \}, \alpha_\gamma \} = 0.
\end{equation}
The equation of motion for
$\alpha_\gamma$ and $\alpha_\gamma^{\ast}$ 
are respectively
\begin{subequations}
\begin{align}
& 
\dot{\alpha}_\gamma (t)
= -i\omega_0 \alpha_\gamma (t) \left[ 
	1 + \frac{\gamma \sigma_0}{\sqrt{2}} 
	(\alpha_\gamma^{\ast} (t) + \alpha_\gamma (t)) 
	\right],
\\
& \dot{\alpha}_\gamma^{\ast} (t)
= i\omega_0 \alpha_\gamma^{\ast} (t) \left[ 
	1 + \frac{\gamma \sigma_0}{\sqrt{2}} 
	(\alpha_\gamma^{\ast} (t) + \alpha_\gamma (t)) 
	\right].
\end{align}
\end{subequations}

Considering the ansatz 
$\alpha_\gamma(t) = |\alpha_\gamma| e^{-i\theta_\gamma (t)}$,
we get $\dot{\theta}_\gamma(t) = \omega_0 (1+\gamma x(t))$.
From 
\begin{equation}
x(t) = \sqrt{\frac{\hbar}{2m_0 \omega_0}} 
		 [{\alpha}_{\gamma}^{\ast}(t) + {\alpha}_{\gamma}(t)] 
	 = A \cos \theta_\gamma (t)
\end{equation}
with amplitude $A = \sqrt{2} \sigma_0 |\alpha_\gamma|$,
we can write
\begin{equation}
\int^{\theta_\gamma} \frac{\textrm{d}\theta}{1+\gamma A \cos \theta} 
	= \omega_0 (t-t_0),
	\qquad (\theta_\gamma (t_0) = 0).
\end{equation}
Consequently, the deformed phase is
\begin{equation}
\label{eq:deformed-phase}
\theta_\gamma (t) = 
	2\textrm{tan}^{-1} \left\{ 
	\sqrt {\frac{1 + \gamma A}{1 - \gamma A}}\textrm{tan} \left[ 
	\frac{1}{2} \Omega_\gamma (t-t_0)
	\right] \right\},
\end{equation}
where 
$\Omega_\gamma = \sqrt{1-\gamma ^2 A ^2} \omega_0$ 
($\gamma^2 A^2 < 1$)
is the angular frequency of the oscillator.
The linear momentum of 
the oscillator evolves according to
\begin{equation}
p(t) = - m_0 \omega_0 A  
		\left[ 
			\frac{\sin \theta_\gamma (t)}{ 1 + \gamma A \cos \theta_\gamma (t)} 
		\right].
\end{equation}
Since $A^2 = 2E/m_0 \omega_0^2$ with $\mathcal{H} = E$ 
the energy of the oscillator, 
$\Omega_\gamma$ results dependent 
on the energy of the system for $\gamma \neq 0$.
For $\gamma^2 A^2 > 1$, the phase becomes a hyperbolic tangent function and
the system looses its oscillatory behavior
(see Ref.~\onlinecite{Costa-Borges-2018} for more details).

By means of the canonical transformation
(\ref{eq:classical-dynamical-variables}) 
the Hamiltonian (\ref{eq:hamiltonian_oscillator}) 
is mapped into a Morse oscillator
\cite{CostaFilho-Alencar-Skagerstam-AndradeJr-2013,Costa-Borges-2018}
\begin{equation}
\mathcal{K} (x_{\gamma}, \Pi_{\gamma}) 
			= \frac{1}{2m_0}\Pi_\gamma^2 
			+ W_\gamma (e^{-\kappa_\gamma x_\gamma} - 1)^2,
\end{equation}
with the binding energy $W_\gamma = m_0 \omega_0^2/2\gamma^2$
and the parameter of anamorticity $\kappa_\gamma = -\gamma$.
Thus, we have that the factorization method applied to 
the oscillator with PDM also allows to obtain the solutions 
of the classic Morse oscillator, which
\begin{subequations}
\begin{align}
 x_\gamma (t) &= \frac{1}{\gamma} 
				\ln (1 + \gamma A \cos \theta_\gamma (t)),  \\
 \Pi_\gamma (t) &= -m_0 \omega_0 A \sin \theta_\gamma (t)
\end{align}
\end{subequations}
and $\gamma^2 A^2 = E/W_\gamma$.
Paths in both representations 
$(x, p)$ and $(x_\gamma, \Pi_\gamma)$
can be found in Ref.~\onlinecite{Costa-Borges-2018}.

\begin{figure}[!htb]
\centering
\includegraphics[width=0.38\linewidth]{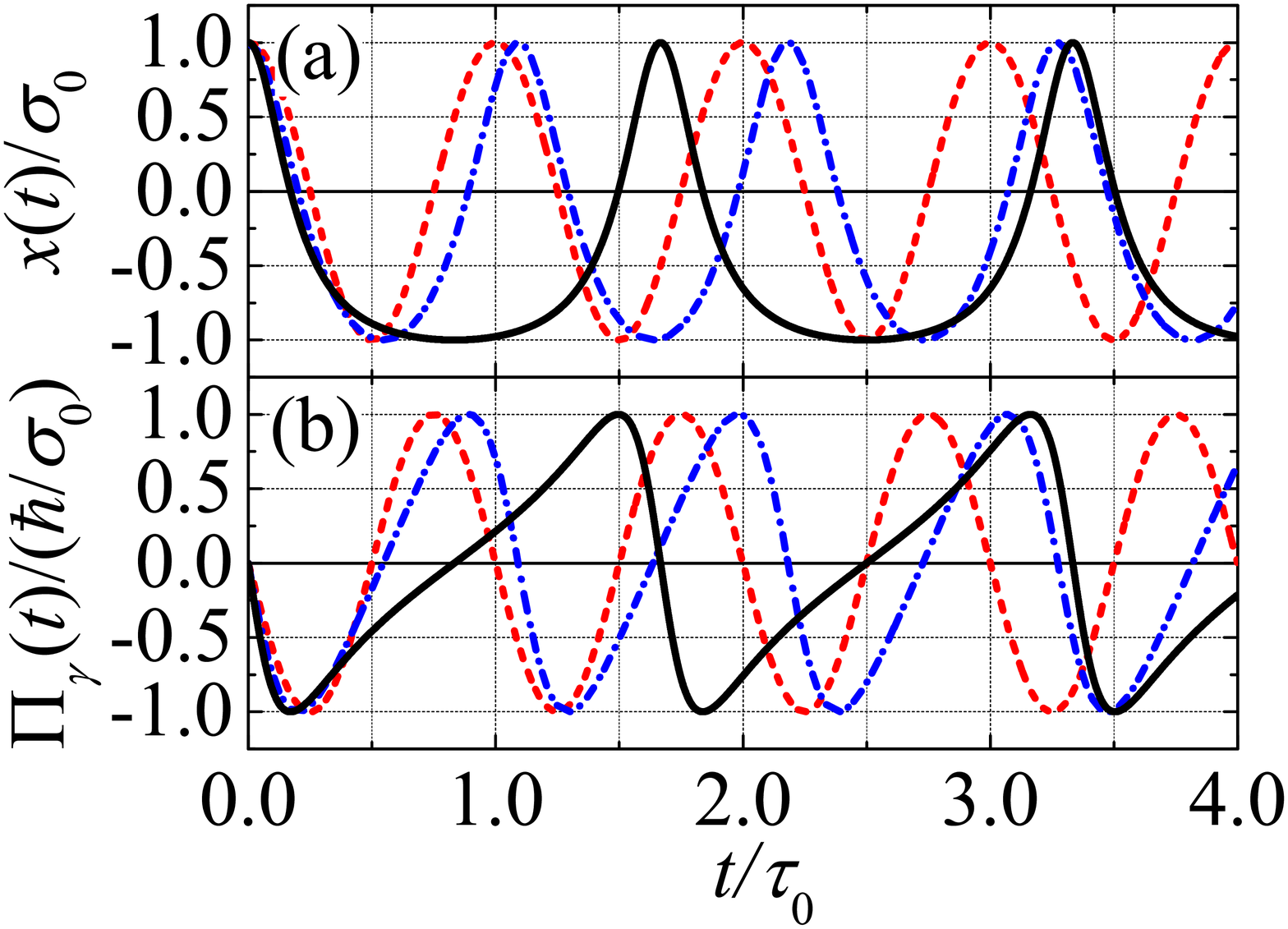}
\includegraphics[width=0.30\linewidth]{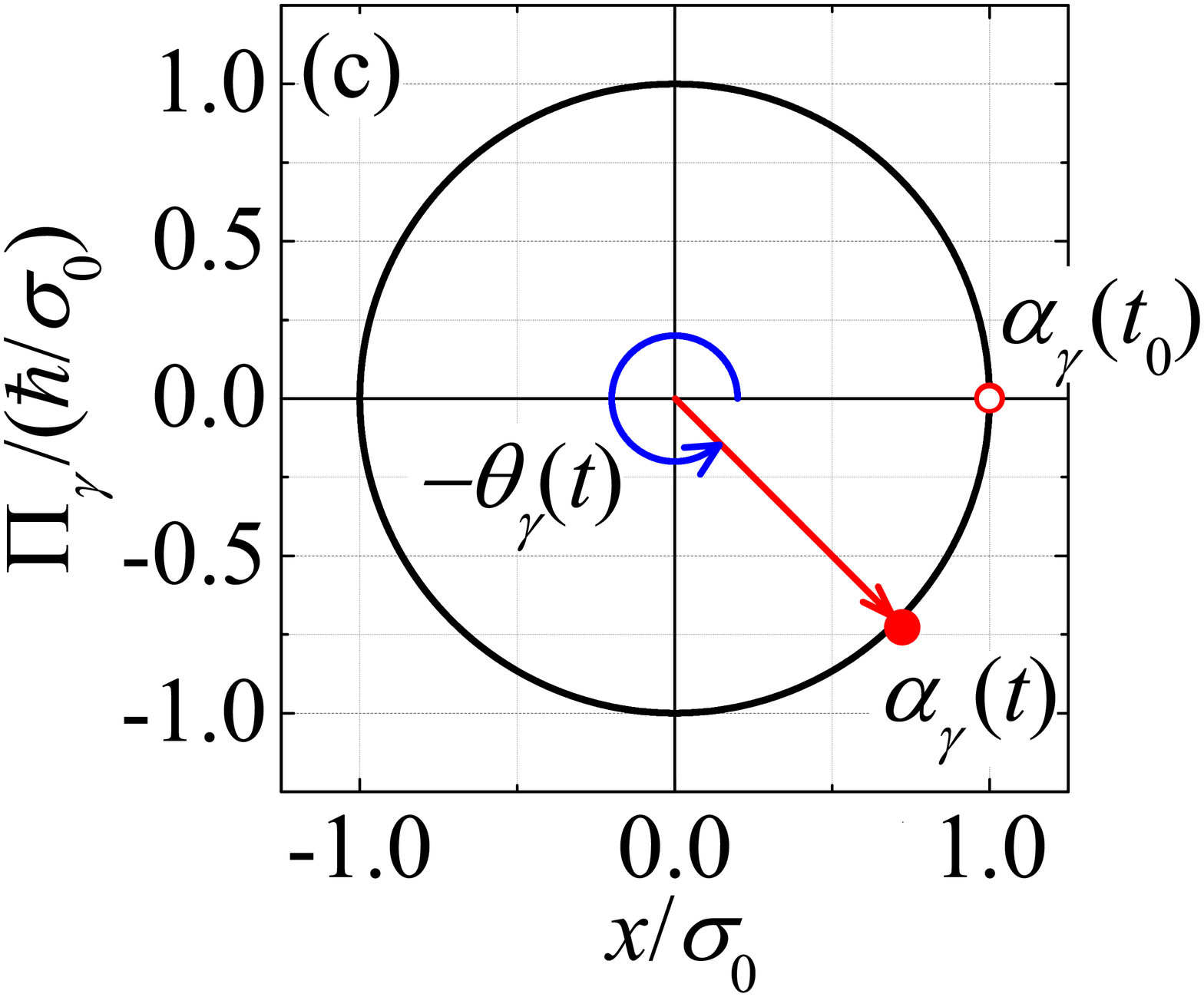}
\caption{\label{fig:1}
Temporal evolution of (a) position $x(t)$ and 
(b) linear pseudo-momentum $\Pi_\gamma (t)$
for a PDM classical oscillator with 
deformed phase $\theta_\gamma(t)$ given by Eq.~(\ref{eq:deformed-phase}) 
for $|\alpha_\gamma| = 1/\sqrt{2}$ ($A=\sigma_0$) 
and deformation parameters
$\gamma \sigma_0 = 0$ (dashed red),
$0.4$ (dashed-dotted blue) and
$0.8$ (solid black),
(with $\tau_0 = 2\pi/\omega_0$ and $t_0=0$).
(c) 
Complex number $\alpha_\gamma$, which corresponds to the state 
of the deformed oscillator,
moves in an elliptic path.
The abscissa and ordinate of $\alpha_\gamma(t)$ 
give $x(t)$ and $\Pi_\gamma(t)$, respectively. 
}
\end{figure}

The point $(x(t), \Pi_\gamma(t))$, 
which corresponds to the complex number $\alpha_\gamma(t)$, 
describes an elliptic path with (nonlinear) phase oscillation 
$\theta_\gamma(t)$ [Eq.~(\ref{eq:deformed-phase})],
since
$\mathcal{H}(x,p) = 
\frac{1}{2m_0}\Pi_\gamma^2(x,p) 
+ \frac{1}{2}m_0\omega_0^2 x^2 = E$.
In terms of the deformed time derivative 
$\mathcal{D}_\gamma (\,\cdot\,)
= \frac{1}{1+\gamma x} \frac{\textrm{d}}{\textrm{d}t}(\,\cdot\,)$,
the classical equations of motion can be conveniently written as
\begin{equation}
\label{eq:classical_equation_of_motion}
\left\{
\begin{array}{rcl}
\displaystyle
\frac{1}{\sigma_0} \displaystyle \mathcal{D}_\gamma x(t)  &=& 
\displaystyle
\frac{\sigma_0}{\hbar} \Pi_\gamma (t), \\
\displaystyle
\frac{\sigma_0}{\hbar} \displaystyle \mathcal{D}_\gamma \Pi_\gamma (t)  &=& 
\displaystyle
- \frac{1}{\sigma_0}x(t).
\end{array}
\right.
\end{equation} 
Figure \ref{fig:1} shows the effect of 
the deformation parameter $\gamma$ on the oscillation phase
(\ref{eq:deformed-phase}), represented in the 
time evolution of the position $x(t)$ and 
linear pseudo-momentum $\Pi_\gamma(t)$.

\subsection{Quantum oscillator with PDM}

Considering the problem of the oscillator with PDM given by (\ref{eq:m(x)})
under the quadratic potential $V(x) = \frac{1}{2}m_0\omega_0^2 x^2,$
the deformed Schr\"odinger equation is
\cite{CostaFilho-Alencar-Skagerstam-AndradeJr-2013,Costa-Borges-2018}
\begin{equation}
\label{eq:deformed-schrodinger-equation-and-derivative}
-\frac{\hbar^2}{2m_0} D_{\gamma}^2 \varphi (x)
+ \frac{1}{2} m_0 \omega_0^2 x^2 
 \varphi (x) =  E \varphi (x).
\end{equation}
In the deformed space $x_\gamma$,
the Eq.~(\ref{eq:deformed-schrodinger-equation-and-derivative})
can be expressed in terms of a new field 
$\phi(x_\gamma) = \varphi(x(x_\gamma))$
and it becomes the Schr\"odinger equation 
for the quantum Morse oscillator
\cite{CostaFilho-Alencar-Skagerstam-AndradeJr-2013,
	  Costa-Borges-2018}
\begin{equation}
\label{eq:SE-QMO}
-\frac{\hbar^2}{2m_0} \frac{\textrm{d}^2{\phi}(x_\gamma)}{\textrm{d}x_\gamma^2}
+W_\gamma (e^{-\kappa_\gamma x_\gamma}-1)^2 \phi ({x_\gamma})
= E \phi ({x_\gamma}).
\end{equation}
From the solutions of (\ref{eq:SE-QMO}), 
we obtain the normalized eigenfunctions 
of the deformed oscillator 
\begin{align}
\label{eq:egeinfucntions-osc}
\psi_n (x)  &= \frac{\varphi_n (x)}{\sqrt{1+\gamma x}} \nonumber \\
			&= (-1)^n  \mathcal{N}_n \sqrt{2s} e^{-\frac{z(x)}{2}} 
	          [z(x)]^{\frac{\nu_n-1}{2}} L_n^{(\nu_{n})} (z(x))
\end{align}
for $x > -1/\gamma$ and $\psi_n (x)=0$  otherwise,
where $z(x) = 2s(1+\gamma x)$, $s = 1/\gamma^2 \sigma_0^2$,
$\nu_n=2s - 2n - 1>0$,
$\mathcal{N}_n^2 = \nu_n \gamma \Gamma(n+1)/\Gamma(\nu_n+n+1)$ 
and
$L_n^{(\nu)}(z)$ are the associated Laguerre polynomials.
The energy eigenvalues are
\begin{equation}
\label{eq:egeinvalues-osc-elec-field}
E_n({\gamma}) = \hbar \omega_0 \left( n+\frac{1}{2} \right)
	- \frac{\hbar^2 \gamma^2}{2m_0} \left( n+\frac{1}{2} \right)^2.
\end{equation}

In particular, the eigenfunction for $n=0$ is
\begin{equation}
\label{eq:ground-state}
\psi_0 (x) 	= \sqrt{ \frac{2}{\gamma \sigma_0^2 
				\Gamma \left( \frac{2}{\gamma^2 \sigma_0^2}-1 \right)} }
			e^{-\frac{z(x)}{2}} [z(x)]^{\frac{1}{\gamma^2 \sigma_0^2} - 1}.
\end{equation}
Thus, instead of the standard case, 
whose ground state is a Gaussian package,
the probability density 
$\rho_0 (x) = |\psi_0(x)|^2$
behaves like a Gamma distribution
\begin{equation}
\label{eq:Gamma_distribuition}
\rho_0 (x) = 
	\frac{2}{\gamma \sigma_0^2 \Gamma (\lambda)}
	e^{-z(x)} [z(x)]^{\lambda-1},
\end{equation}
with $\lambda = \frac{2}{\gamma^2 \sigma_0^2} -1$ 
being the shape parameter.
The ground state energy is
$
E_0 ({\gamma}) = \frac{\hbar \omega_0}{2} 
                 - \frac{\hbar^2 \gamma^2}{8m_0}.
$

It is straightforward to verify that
the expected values 
$ \langle \hat{x} \rangle $,
$ \langle \hat{x}^2 \rangle $,
$ \langle \hat{\Pi}_\gamma \rangle $ and
$ \langle \hat{\Pi}_\gamma^2 \rangle $
for the $n$th state are given by
\cite{CostaFilho-Alencar-Skagerstam-AndradeJr-2013,
	  Costa-Borges-2018}
\begin{subequations}
\label{eq:x-p-x^2-p^2-expec-values}
\begin{align}
\label{eq:expected-value-x-oscillator}
& \langle \hat{x} \rangle 
= - \gamma \sigma_0^2 \left( n + \frac{1}{2} \right), \\
\label{eq:expected-value-x^2-oscillator}
& \langle \hat{x}^2 \rangle 
= \sigma_0^2  \left( n + \frac{1}{2} \right), \\
\label{eq:expected-value-p-oscillator}
& \langle \hat{\Pi}_\gamma \rangle = 0, \\
& \label{eq:expected-value-p^2-oscillator}
 \langle \hat{\Pi}_\gamma^2 \rangle 
= \frac{\hbar^2}{\sigma_0^2} \left( n + \frac{1}{2} \right)
		\left[ 1 - \gamma^2 \sigma_0^2 
		\left( n + \frac{1}{2} \right) \right].
\end{align}
\end{subequations}
The product of the uncertainties
of the observables $\hat{x}$ and $\hat{\Pi}_\gamma$ is 
\begin{align}
\label{eq:DxDPi_gamma}
\Delta x \Delta \Pi_\gamma
 		&= 
		\hbar \left( n + \frac{1}{2} \right)
		  \left[ 
			1 - \gamma^2 \sigma_0^2 \left( n + \frac{1}{2} \right)
			\right]
		\nonumber \\
		&= \hbar \left( n + \frac{1}{2} \right)
		  (1 + \gamma \langle \hat{x} \rangle ),
\end{align}
which obeys the deformed uncertainty relation
(\ref{eq:uncertainty-relation}).

\subsection{Supersymmetric quantum mechanics for deformed oscillator}

The Hamiltonian operator (\ref{eq:hamiltonian-MM}) for the problem 
of a quadratic potential 
$V(\hat{x}) = \frac{1}{2}m_0\omega_0 \hat{x}^2$
can be factorized as
\begin{equation}
\label{eq:H-factorized}
\hat{H} = \hbar \omega_0 \hat{a}_\gamma^{\dagger} \hat{a}_\gamma
		  + E_0 ({\gamma}), 
\end{equation}
with the annihilation and the creation operators given by
\begin{subequations}
\label{eq:a_gamma-and-a_gamma-dagger}
\begin{align}
\label{eq:a_gamma}
\hat{a}_\gamma 
	&= \sqrt{\frac{m_0 \omega_0}{2\hbar}}	
		\left[ 
		\hat{x} + \frac{ \hbar \gamma}{m_0 \omega_0} \hat{1} 
		+ \frac{i}{m_0\omega_0} (\hat{1} + \gamma \hat{x}) \hat{p} 
		\right] 
		\nonumber \\
	&= \sqrt{\frac{m_0 \omega_0}{2\hbar}}	
		\left(
		\hat{x} + \frac{ \hbar \gamma}{2m_0 \omega_0} \hat{1} 
		+ \frac{i}{m_0\omega_0} \hat{\Pi}_\gamma \right) 
\end{align}
and
\begin{align}
\label{eq:a_gamma_dagger}
\hat{a}_\gamma^{\dagger}
	&= \sqrt{\frac{m_0 \omega_0}{2\hbar}}	
		\left[ \hat{x} + \frac{ \hbar \gamma}{m_0 \omega_0} \hat{1} 
		- \frac{i}{m_0\omega_0} 
		\hat{p}(\hat{1} + \gamma \hat{x}) 
		\right]
		\nonumber \\
	&= \sqrt{\frac{m_0 \omega_0}{2\hbar}}	
		\left( \hat{x} + \frac{ \hbar \gamma}{2m_0 \omega_0} \hat{1} 
		- \frac{i}{m_0\omega_0} \hat{\Pi}_\gamma 
		\right).
\end{align}
\end{subequations}
We see that $\hat{a}_\gamma \psi_0(x) = 0$, i.e.,
the ground state is annihilated by the operator $\hat{a}_\gamma$.
Although $\hat{a}_\gamma$ and $\hat{a}_\gamma^{\dagger}$
factorize the Hamiltonian (\ref{eq:hamiltonian-MM})
for the quadratic potential,
these not are the ladder operators 
since they satisfy the commutator
$
[\hat{a}_\gamma, \hat{a}_\gamma^{\dagger}] = \hat{1}+\gamma \hat{x}.
$
A deformed number operator 
$
 \hat{n}_\gamma = \hat{a}_\gamma^{\dagger} \hat{a}_\gamma
$ 
leads to the expected value for $n$th state
$
 \langle \hat{n}_\gamma \rangle = 
 n\left[ 1 - \frac{1}{2} \gamma^2 \sigma_0^2 (n+1) \right]
\equiv [n_{\gamma}].
$

We can straightforwardly verify that the operators
(\ref{eq:a_gamma}) and (\ref{eq:a_gamma_dagger})
satisfy the commutation relations
\begin{subequations}
\begin{align}
[\hat{a}_\gamma, \hat{a}_\gamma^{\dagger} \hat{a}_\gamma ]
	&= (\hat{1} +\gamma \hat{x}) \hat{a}_\gamma
	\nonumber \\
	&= \left( 1 - \gamma^2 \sigma_0^2 + \frac{\gamma \sigma_0}{\sqrt{2}} 
	(\hat{a}_\gamma^{\dagger} + \hat{a}_{\gamma}) \right) \hat{a}_\gamma,
	\\
[\hat{a}_\gamma^{\dagger}, \hat{a}_\gamma^{\dagger} \hat{a}_\gamma ]
	&= -\hat{a}_\gamma^{\dagger}(\hat{1} +\gamma \hat{x}) 
	\nonumber \\
	&= -\hat{a}_\gamma^{\dagger}
	\left( 1 - \gamma^2 \sigma_0^2 + \frac{\gamma \sigma_0}{\sqrt{2}} 
	( \hat{a}_\gamma^{\dagger} + \hat{a}_{\gamma})
	\right).
\end{align}
\end{subequations}
Similar to Eq.~(\ref{eq:classical-Jacobi}) for the classical analog,  
the Jacobi identify 
\begin{equation}
[[\hat{a}_\gamma, \hat{a}_\gamma^{\dagger}], \hat{H}]+
[[\hat{H}, \hat{a}_\gamma ],\hat{a}_\gamma^{\dagger}]+
[[\hat{a}_\gamma^{\dagger}, \hat{H}],\hat{a}_\gamma] =0
\end{equation}
is satisfied in quantum formalism, so that 
the operators $\hat{a}_\gamma$, $\hat{a}_\gamma^{\dagger}$
and $\hat{H}$ constitute a Lie algebra.

The supersymmetric partners Hamiltonian operators 
\cite{Plastino-etal-1999,Amir-Iqbal-2016}
associated with $\hat{H}$ are
\begin{subequations}
\begin{equation}
\label{eq:H_+} 
	\hat{H}_{\scriptscriptstyle +} 
	= \hbar \omega_0 \hat{a}_{\gamma}^{\dagger} \hat{a}_\gamma 
	= \frac{1}{2m_0} \hat{\Pi}_\gamma^2 + 
	V_{\scriptscriptstyle +}(\hat{x})
\end{equation}
and
\begin{equation}
\label{eq:H_-}
	\hat{H}_{\scriptscriptstyle -} 
	= \hbar \omega_0 \hat{a}_\gamma \hat{a}_{\gamma}^{\dagger}  
	= \frac{1}{2m_0} \hat{\Pi}_\gamma^2 + 
	V_{\scriptscriptstyle -}(\hat{x}),
\end{equation}
\end{subequations}
whose potentials are 
$
V_{\scriptscriptstyle +} (\hat{x}) = 
\frac{1}{2}m_0\omega_0^2 \hat{x}^2 - E_0(\gamma)
$
and
$
V_{\scriptscriptstyle -} (\hat{x}) = 
\frac{1}{2}m_0\omega_0^2 \hat{x}^2 - E_0(\gamma)
+ \hbar \omega_0 (\hat{1} + \gamma \hat{x}),
$
respectively.
That is, the potential of the partner operator 
$\hat{H}_{\scriptscriptstyle +} $ has only 
a shift equal to the energy of the ground state
in relation to the original operator $\hat{H}$, 
while the potential of the partner operator 
$\hat{H}_{\scriptscriptstyle -}$ 
has also added a quanta of energy, 
$\hbar \omega_0$, and a term corresponding to 
a uniform electric field, 
$\hbar \omega_0 \gamma \hat{x}$.

Denoting the eigenenergies and eigenfunctions equations of 
the partners operators as
$\hat{H}_{\scriptscriptstyle +} \psi_n^{\scriptscriptstyle (+)}(x) 
= E_n^{\scriptscriptstyle (+)}\psi_n^{\scriptscriptstyle (+)}(x)$
and 
$\hat{H}_{\scriptscriptstyle -} \psi_n^{\scriptscriptstyle (-)}(x) 
= E_n^{\scriptscriptstyle (-)}\psi_n^{\scriptscriptstyle (-)}(x)$.
Since $\hat{H}_{\scriptscriptstyle +}$ and $\hat{H}$ commute, then 
$\psi_n^{\scriptscriptstyle (+)}(x) = \psi_n (x)$.
In addition, from wave function
$
 \varphi_n^{\scriptscriptstyle (-)}(x) = 
 \sqrt{1+\gamma x} \psi_n^{\scriptscriptstyle (-)}(x)
$
and 
$
 \epsilon_n = E_{n}^{\scriptscriptstyle (-)} 
			  - \hbar \omega_0 + E_0 (\gamma),
$
we arrive at the deformed Schr\"odinger equation
\begin{equation}
\label{eq:deformed-SE-electric-field}
	-\frac{\hbar^2}{2m_0} D_{\gamma}^2 \varphi_n^{\scriptscriptstyle (-)} (x)
	+ \frac{1}{2} m_0 \omega_0  x^2 \varphi_n^{\scriptscriptstyle (-)} (x) 
	+ \hbar \omega_0 \gamma x \varphi_n^{\scriptscriptstyle (-)} (x)
	= \epsilon_n \varphi_n^{\scriptscriptstyle (-)} (x).
\end{equation}
From the change of variable $x \rightarrow x_\gamma$,
Eq.~(\ref{eq:deformed-SE-electric-field}) becomes also
a Morse oscillator
\begin{equation}
\label{eq:deformed-SE-electric-field2}
-\frac{\hbar^2}{2m_0} 
\frac{\textrm{d}^2 \phi_n^{\scriptscriptstyle (-)}(x_\gamma)}{\textrm{d} x_\gamma^2}
+ \widetilde{W}_\gamma  (e^{-\kappa_\gamma (x_\gamma - \delta_\gamma)} - 1)^2 
\phi_n^{\scriptscriptstyle (-)} (x_\gamma) 
= \left( \epsilon_n + \frac{\hbar^2 \gamma^2}{2m_0} \right) 
    \phi_n^{\scriptscriptstyle (-)} (x_\gamma),
\end{equation}
where 
$
\phi_n^{\scriptscriptstyle (-)} (x_\gamma) =
\varphi_n^{\scriptscriptstyle (-)} (x(x_\gamma)),
$
$\widetilde{W}_\gamma = m_0 \omega_\gamma^2/2\gamma^2$ 
is a shifted binding energy, and
$\omega_\gamma = \omega_0 (1 - \gamma^2 \sigma_0^2)$ is
the frequency of small oscillations around the equilibrium position
$\delta_\gamma = \gamma^{-1} \ln (1 - \gamma^2 \sigma_0^2)$.
The solution of the above equations leads
to the eigenfunctions
\begin{align}
\label{eq:psi_n^-} 
\psi_{n}^{\scriptscriptstyle (-)}(x) &=
    \frac{\varphi_n^{\scriptscriptstyle (-)} (x)}{\sqrt{1+\gamma x}} \nonumber \\
    &=(-1)^{n}\sqrt{2s} \widetilde{\mathcal{N}}_n e^{-\frac{z(x)}{2}} 
	[z(x)]^{\frac{\widetilde{\nu}_{n}-1}{2}} 
	L_n^{(\widetilde{\nu}_n)} (z(x)),
\end{align}
with 
$\widetilde{\nu}_n = 2s-2n-3 $ 
and
$
\widetilde{\mathcal{N}}_n^2 = 
  \widetilde{\nu}_n\gamma \Gamma(n+1)/\Gamma(\widetilde{\nu}_n+n+1).
$

The correspondence between the partner operators 
is established as follows
\begin{subequations}
\begin{equation}
\hat{H}_{\scriptscriptstyle -} (\hat{a}_\gamma \psi_n^{\scriptscriptstyle (+)}) 
= \hat{a}_\gamma (\hat{H}_{\scriptscriptstyle +} \psi_n^{\scriptscriptstyle (+)})
= E_n^{\scriptscriptstyle (+)} (\hat{a}_\gamma \psi_n^{\scriptscriptstyle (+)})
\end{equation}
and
\begin{equation}
\hat{H}_{\scriptscriptstyle +} 
(\hat{a}_\gamma^{\dagger} \psi_n^{\scriptscriptstyle (-)}) 
= \hat{a}_\gamma^{\dagger} (\hat{H}_{\scriptscriptstyle -} 
\psi_n^{\scriptscriptstyle (-)})
= E_n^{\scriptscriptstyle (-)} (\hat{a}_\gamma^{\dagger} 
\psi_n^{\scriptscriptstyle (-)}),
\end{equation}
\end{subequations}
so that, $\hat{a}_\gamma \psi_n^{\scriptscriptstyle (+)}$ 
($\hat{a}_\gamma^{\dagger} \psi_n^{\scriptscriptstyle (-)}$)
is eigenfunction of $\hat{H}_{\scriptscriptstyle -}$ 
($\hat{H}_{\scriptscriptstyle +}$).
The energy spectra of the operators 
(\ref{eq:H_+}) and (\ref{eq:H_-}) 
obey the recurrence relation
\begin{equation}
\label{eq:E_n^+andE_n^-}
E_{n}^{\scriptscriptstyle (+)} 
		  = E_{n-1}^{\scriptscriptstyle (-)} 
		  = E_n - E_0({\gamma})
		  = \hbar \omega_0 [n_{\gamma}]
\end{equation}
and $E_0^{\scriptscriptstyle (+)}=0$. 
The eigenfunctions are related as
\begin{subequations}
\label{eq:a_gamma-and-adjunct-eigenfunctions}
\begin{equation}
\label{eq:a_gamma-eigenfunctions}
\hat{a}_\gamma \psi_n^{\scriptscriptstyle (+)}(x)
	= \left[\frac{E_n^{\scriptscriptstyle (+)}}{\hbar \omega_0}\right]^{1/2} 
	  \psi_{n-1}^{\scriptscriptstyle (-)} (x)
\end{equation}
and
\begin{equation}
\label{eq:a_gamma-dagger-eigenfunctions}
\hat{a}_\gamma^{\dagger} \psi_n^{\scriptscriptstyle (-)}(x)
	= \left[\frac{E_n^{\scriptscriptstyle (-)}}{\hbar \omega_0}\right]^{1/2}
	   \psi_{n+1}^{\scriptscriptstyle (+)} (x),
\end{equation}
\end{subequations}
in which can be verified
from Eqs.~(\ref{eq:a_gamma-and-a_gamma-dagger}), (\ref{eq:E_n^+andE_n^-})
and the expressions of $\psi_n^{\scriptscriptstyle (+)}(x)$ 
and $\psi_n^{\scriptscriptstyle (-)}(x)$
(see Appendix for more details).

The operators (\ref{eq:a_gamma-and-a_gamma-dagger}) 
are not the bosonic operators since 
\begin{equation}
\hat{H} = \frac{\hbar \omega_0}{2} 
\{ \hat{a}_\gamma,\hat{a}_\gamma^{\dagger} \}
-\frac{1}{2}\hbar \omega_0 (\hat{1}+ \gamma \hat{x}) + E_0(\gamma).
\end{equation}
However, the operators (\ref{eq:a_gamma-and-a_gamma-dagger})
can be used to obtain the supersymmetric Hamiltonian operator
associated to the oscillator with PDM. 
In fact, considering the deformed bosonic operator
$\hat{b}_\gamma 
= \hat{a}_\gamma - \frac{\gamma \sigma_0}{2\sqrt{2}} \hat{1}
= \sqrt{\frac{m_0 \omega_0}{2\hbar}} 
\left(\hat{x} + \frac{i}{m_0 \omega_0} \hat{\Pi}_\gamma \right)$
and its adjoint $\hat{b}_\gamma^{\dagger}$, 
we get $\hat{H}$ expressed in terms of an anticommutator, 
\begin{equation}
\label{eq:bosonic-Hamiltonian}
\hat{H} = \frac{\hbar \omega_0}{2} 
\{ \hat{b}_\gamma , \hat{b}_\gamma^{\dagger} \}.
\end{equation}
The Hamiltonian (\ref{eq:bosonic-Hamiltonian})
can be rewritten in terms of the 
deformed bosonic operators as
\begin{equation}
\label{eq:H-bosonic-operators}
\hat{H} =\hbar \omega_0
\left( \hat{b}_\gamma^{\dagger} \hat{b}_\gamma + \frac{1}{2} \right)
+ \frac{\hbar \omega_0 \gamma \sigma_0 }{2\sqrt{2}} 
( \hat{b}_\gamma^{\dagger}  + \hat{b}_\gamma ).
\end{equation}
The first term in (\ref{eq:H-bosonic-operators}) 
corresponds to the Hamiltonian operator
of a deformed oscillator, i.e., 
$\hat{h}_\gamma = \hbar \omega_0 
\left( \hat{b}_\gamma^{\dagger} \hat{b}_\gamma 
+ \frac{1}{2} \right)$,
while the second term is equivalent to an interaction potential 
of a uniform electric field 
due to the effect of the effective mass, 
$\hat{\mathcal{V}}_\gamma = \frac{1}{2}\hbar \omega_0 \gamma \hat{x}$.
Therefore, the supersymmetric Hamiltonian is
\begin{align}
\hat{H}_{\textrm{ss}} 
&= \hbar \omega_0 \left(
\begin{array}{cc}
\hat{b}_\gamma^{\dagger}\hat{b}_\gamma & 0\\
0 & \hat{b}_\gamma \hat{b}_\gamma^{\dagger}	
\end{array}
\right) \nonumber \\
&=
\left(
\begin{array}{cc}
\hat{H} -\frac{\hbar \omega_0}{2}(\hat{1}+ \gamma \hat{x}) & 0\\
0 & \hat{H} + \frac{\hbar \omega_0}{2}(\hat{1} + \gamma \hat{x})
\end{array}
\right),
\end{align}
or more compactly,
$
\hat{H}_{\textrm{ss}} = \hat{H}
- \left(\frac{\hbar \omega_0}{2} 
+ \hat{\mathcal{V}}_\gamma \right) \widehat{\sigma}_z,
$
where $\widehat{\sigma}_z$ is 
the diagonal Pauli matrix.

\subsection{Shape invariance}

For the sake of completeness, let us consider the {\it shape invariance} 
technique to determine the wave functions and the energy spectrum
for the deformed oscillator with PDM.
The partner Hamiltonians are named shape invariants if 
they satisfy the integrability condition 
\cite{Plastino-etal-1999,Amir-Iqbal-2016}
\begin{equation}
\label{eq:integrability-condition}
	\hat{H}_{\scriptscriptstyle -} ({\beta}_{j})
	- \hat{H}_{\scriptscriptstyle +}({\beta}_{j+1}) 
	= R({\beta}_{j}),
\end{equation}
so that the set of parameters is 
related by a function $f$ such that 
$\beta_{j+1} = f(\beta_j)$, 
and the remainder term $R(\beta_j)$ is independent 
of the position and linear momentum operators.
Since the partner operators differ only by an additive constant, 
their energy spectra and eigenstates are related respectively as
\begin{subequations}
\label{eq:energy_and_states_SI}
\begin{align}	
	E_n^{\scriptscriptstyle (-)} ({\beta}_{j})
	& = E_n^{\scriptscriptstyle (+)}({\beta}_{j+1}) 
	+ R({\beta}_{j}), \\
	|\psi_n^{\scriptscriptstyle (-)} (\beta_{j}) \rangle 
	& = |\psi_n^{\scriptscriptstyle (+)} (\beta_{j+1}) \rangle
	  = |\psi_n (\beta_{j+1}) \rangle.
\end{align}
\end{subequations}

The application of the shape invariance method to 
the deformed oscillator leads to change 
the intertwining operators
(\ref{eq:a_gamma-and-a_gamma-dagger}) so that
\begin{subequations}
\label{eq:a_gamma-and-a_gamma-dagger_beta}
\begin{align}
\label{eq:a_gamma_beta}
\hat{a}_\gamma ({\beta})
	&= \sqrt{\frac{m_0 \omega_0}{2\hbar}}	
		\left( 
		\hat{x} + \frac{ \beta \hbar \gamma}{2m_0 \omega_0} \hat{1} 
		+ \frac{i}{m_0\omega_0} \hat{\Pi}_\gamma 
		\right)
\end{align}
and
\begin{align}
\label{eq:a_gamma_dagger_beta}
\hat{a}_\gamma^{\dagger} ({\beta})
	&= \sqrt{\frac{m_0 \omega_0}{2\hbar}}	
		\left( \hat{x} + \frac{ \beta \hbar \gamma}{2m_0 \omega_0} \hat{1} 
		- \frac{i}{m_0\omega_0} \hat{\Pi}_\gamma 
		\right).
\end{align}
\end{subequations}
The creation and annihilation operators (\ref{eq:a_gamma-and-a_gamma-dagger})
are recovered as $\beta \rightarrow 1$.
The supersymmetric partner Hamiltonians 
$
\hat{H}_{\scriptscriptstyle +} ({\beta}) =
\hbar \omega_0 \hat{a}_\gamma^{\dagger} ({\beta}) \hat{a}_\gamma ({\beta})
$
and
$
\hat{H}_{\scriptscriptstyle -} ({\beta}) = 
\hbar \omega_0 \hat{a}_\gamma ({\beta}) \hat{a}_\gamma^{\dagger} ({\beta}) 
$
are
$
\hat{H}_{\scriptscriptstyle \pm}  = 
\frac{1}{2m_0}\hat{\Pi}_\gamma^2 + V_{\scriptscriptstyle \pm}(\hat{x}, \beta )
$
with potentials 
$
V_{\scriptscriptstyle \pm}(\hat{x}, \beta ) =  V(\hat{x})
\mp\frac{\hbar \omega_0}{2}[\hat{1}+(1 \mp \beta)\gamma \hat{x}]
+\frac{\beta^2 \hbar^2 \gamma^2}{8m_0}.
$

Consequently, the integrability condition 
(\ref{eq:integrability-condition}) becomes 
\begin{equation}
\hbar \omega_0 \hat{a}_\gamma ({\beta}_j) \hat{a}_\gamma^{\dagger} ({\beta}_j)
- \hbar \omega_0 \hat{a}_\gamma^{\dagger} ({\beta_{j+1}}) \hat{a}_\gamma({\beta_{j+1}})
= R(\beta_{j}),
\end{equation}
where the $\beta$-parameters satisfy 
a translational shape invariance
$\beta_{j+1} = \beta_j + \eta$ with $\eta=2$,
such that, $\beta_n = \beta + 2(n-1)$
and $\beta_1 \equiv \beta$.
The remainder term is
$ 
R(\beta) = \hbar \omega_0 \left[ 
			1-\frac{1}{2}\gamma^2 \sigma_0^2 (\beta+1) 	
			\right].
$

The energy levels of the operator 
$\hat{H}_{\scriptscriptstyle +}({\beta})$ 
are obtained from
$E_n^{\scriptscriptstyle (+)}({\beta}) 
= \sum_{j=1}^{n} R({\beta}_j),$
with
$E_0^{\scriptscriptstyle (+)}({\beta}) = 0.$
In this way, it is straightforwardly to verify that
\begin{equation}
E_n^{\scriptscriptstyle (+)}({\beta}) = 
		\hbar \omega_0 n 
		\left[ 
			1 - \frac{\gamma^2 \sigma_0^2}{2} ( n+\beta ) 
		\right].
\end{equation}
Thereby, the operator
$\hat{H}(\beta) = \hbar \omega_0 
\hat{a}_\gamma^{\dagger}(\beta) \hat{a}_\gamma(\beta) + E_0(\gamma)$ 
has energy spectrum
$E_n = E_n^{\scriptscriptstyle (+)}({\beta}) + E_0 (\gamma).$

The next step is to obtain the eigenfunctions 
using the shape invariance method.
From Eq.~(\ref{eq:a_gamma-eigenfunctions}), 
the eigenstates of the oscillator satisfy 
the recurrence relation
\begin{equation}
|\psi_n ({\beta}_j)\rangle =
	\left[ 
		\frac{E_n^{\scriptscriptstyle (+)}}{\hbar \omega_0} 
	\right]^{-1/2}	
		\hat{a}_\gamma^{\dagger} ({\beta}_{j+1}) 
		|\psi_{n-1} (\beta_{j+1})\rangle.
\end{equation}
Applying $n$ interactions, it can be seen that
\begin{equation}
\label{eq:psi_n_a_dagger_n}
	|\psi_n (\beta_{1}) \rangle =
	\frac{1}{\sqrt{[n_\gamma ({\beta})]!}} 
	\hat{a}_\gamma^{\dagger} ({\beta}_1)
	\hat{a}_\gamma^{\dagger} ({\beta}_2)
	...
	\hat{a}_\gamma^{\dagger} ({\beta}_n)
	|\psi_0 (\beta_{n+1})\rangle
\end{equation}
with deformed factorial given by
\begin{equation}
\label{eq:[n]!}
[n_\gamma ({\beta})]! 
	= \prod_{j=1}^{n}
	 \left[ 
		\frac{E_j^{\scriptscriptstyle (+)}({\beta})}{\hbar \omega_0} 
	\right] 
	= \frac{n!}{(2s)^n} 
	\frac{\Gamma (2s + 1 - \beta - n)}{\Gamma (2s + 1 - \beta - 2n)}.
\end{equation}

The operator (\ref{eq:a_gamma_dagger_beta}) can be recasted as
\begin{align}
 \hat{a}_\gamma^{\dagger} ({\beta}) y (z)
		&= \frac{1}{\sqrt{2s}}
		\left[
		-\frac{2s - \beta +1}{2} + \frac{z}{2} - z \frac{\textrm{d}}{\textrm{d} z}
		\right] y (z)
		\nonumber \\
		&= -\frac{1}{\sqrt{2s}} 
			\left[ \frac{1}{g_{\scriptscriptstyle \beta}(z)}
			\left( z \frac{\textrm{d}}{\textrm{d}z} \right)
			g_{\scriptscriptstyle \beta}(z) \right] y (z)
\end{align}
with $y(z)$ being a generic function and
$g_{\scriptscriptstyle \beta} (z(x)) = 
e^{-\frac{z(x)}{2}} [z(x)]^{\frac{2s-\beta+1}{2}}.$
From that, we have
\begin{equation}
\label{eq:product_a_dagger}
\prod_{j=1}^n \hat{a}_\gamma^{\dagger} ({\beta}_j) y (z)
	= \left( -\frac{1}{\sqrt{2s}} \right)^{n}
	\left[ \frac{1}{g_{\scriptscriptstyle \beta} (z)} \left(
	z^n \frac{\textrm{d}^n}{\textrm{d}z^n} 
	\right) g_{\scriptscriptstyle \beta} (z)  \right] y (z).
\end{equation}
The condition 
$\hat{a}_\gamma ({\beta}) |\psi_{0} ({\beta}) \rangle = 0$ 
leads to the ground state
\begin{equation}
\label{eq:SI_ground_state}
\psi_{0,\beta} (x) 
		= \sqrt{ \frac{2s\gamma}{\Gamma ( 2s - \beta )} }
		e^{-\frac{z(x)}{2}} [z(x)]^{\frac{2s-1-\beta}{2}}.
\end{equation}
From Eqs.~(\ref{eq:psi_n_a_dagger_n}), (\ref{eq:[n]!}), 
(\ref{eq:product_a_dagger}), (\ref{eq:SI_ground_state}) 
and Rodrigues' formula for 
the associated Laguerre polynomials,
$L_n^{(\nu)}(z) = \frac{1}{n!}e^z z^{-\nu} 
\times \frac{\textrm{d}^n}{\textrm{d}x^n} (z^{\nu+n} e^{-z}),$
we arrive at
\begin{align}
\psi_{n,\beta} (x) 
	&=(-1)^n  \sqrt{2s}\left[ 
			\frac{(\nu_n + 1 - \beta ) \gamma \Gamma (n+1) }{
				  \Gamma (\nu_n + 2 - \beta + n)} 
				\right]^{1/2}
	  e^{-\frac{z(x)}{2}} [z(x)]^{\frac{\nu_n - \beta}{2}}
	  L_n^{( \nu_n + 1 - \beta )}(z(x)).
\end{align}
When the parameter $\beta \rightarrow 1$,
the expression above reduces to the 
eigenfunctions (\ref{eq:egeinfucntions-osc}).

Since the commutator between the $\hat{a}_\gamma(\beta)$ and 
$\hat{a}_\gamma^{\dagger}(\beta)$ depends on the position, 
they can not be chosen as ladder operators.
However, the Hamiltonian operators 
$\hat{H}_{\scriptscriptstyle \pm} ({\beta})$
are translational shape invariance, and in this case 
it is possible to define ladder operators as
\begin{equation}
\label{eq:ladder_operators}
\hat{L}_{\scriptscriptstyle -}({\beta}) = 
\hat{\Lambda}^{\dagger}({\beta})\hat{a}_{\gamma}({\beta})
\quad \textrm{and} \quad
\hat{L}_{\scriptscriptstyle +}({\beta}) =  
\hat{a}_{\gamma}^{\dagger} ({\beta}) \hat{\Lambda}({\beta})
\end{equation}
with $\hat{\Lambda}^{\dagger}({\beta})$ 
and $\hat{\Lambda}({\beta})$ 
unitary translational operators 
on parameter $\beta$, 
which satisfying the reparametrization 
$\hat{\Lambda} (\beta) |\psi_n (\beta)\rangle = 
|\psi_n (\beta + \eta)\rangle.$ 
The translational operators 
$\hat{\Lambda}({\beta})$ and 
$\hat{\Lambda}^{\dagger}({\beta})$ 
are given respectively by 
\cite{Amir-Iqbal-2016}
\begin{equation}
\hat{\Lambda} ({\beta}) = 
	\exp \left( \eta \frac{\partial}{\partial \beta} \right) 
\ \ \textrm{and} \ \
\hat{\Lambda}^{\dagger} ({\beta}) = 
	\exp \left( - \eta \frac{\partial}{\partial \beta} \right). 
\end{equation}

Once $\hat{\Lambda}^{\dagger} (\beta) \hat{\Lambda} (\beta) = \hat{1}$
the Hamiltonian operator (\ref{eq:H_+}) preserves the form
$
\hat{H}_{\scriptscriptstyle +} ({\beta})= \hat{H}(\beta) -E_0(\gamma) =  
\hbar \omega_0 \hat{L}_{\scriptscriptstyle +} ({\beta}) 
\hat{L}_{\scriptscriptstyle -}   ({\beta}).
$
From Eqs.~(\ref{eq:energy_and_states_SI}) and
(\ref{eq:ladder_operators}),  
the action of the ladder operators on the ket vectors 
$|\psi_n\rangle$ is given by 
\begin{subequations}
\label{eq:eq:ladder_operators_C_n}
\begin{align}
&\hat{L}_{\scriptscriptstyle -}  |\psi_n ({\beta})\rangle =
	\sqrt{n \left[ 1-\frac{\gamma^2 \sigma_0^2}{2} ( n+\beta ) \right]}
	|\psi_{n-1} ({\beta})\rangle,
\end{align}
and
\begin{align}
&\hat{L}_{\scriptscriptstyle +} |\psi_n ({\beta})\rangle =
\sqrt{(n+1) \left[ 1-\frac{\gamma^2 \sigma_0^2}{2} ( n+1+\beta ) \right]}
	|\psi_{n+1} ({\beta})\rangle.
\end{align}
\end{subequations}
Explicitly, the ladder operators have the form
\begin{subequations}
\label{eq:eq:ladder_operators_explicitly}
\begin{align}
&\hat{L}_{\scriptscriptstyle -}   ({\beta}) = 
	e^{-\eta \frac{\partial}{\partial \beta}}
	\frac{1}{\sqrt{2} \sigma_0} 
	\left[ 
		x + \frac{(\beta + 1) \gamma \sigma_0^2}{2}
		+ \sigma_0^2 (1 + \gamma x) \frac{\textrm{d}}{\textrm{d}x} 
	\right],
\\
&\hat{L}_{\scriptscriptstyle +}   ({\beta}) = 
	\frac{1}{\sqrt{2} \sigma_0}
	\left[ 
		x + \frac{(\beta - 1) \gamma \sigma_0^2}{2}
		- \sigma_0^2 (1 + \gamma x) \frac{\textrm{d}}{\textrm{d}x} 
	\right] 
	e^{\eta \frac{\partial}{\partial \beta}}.
\end{align}
\end{subequations}
After careful calculations, 
we have found that the effects of the ladder operators on
wavefunctions are expressed as
\begin{align}
\hat{L}_{\scriptscriptstyle \pm}   ({\beta}) \psi_{n,\beta} (x)
=& \frac{1}{\sqrt{2}\sigma_0} 
	\left\{ \frac{1}{\gamma} 
	\left[
		\frac{2s + 1 - \beta }{2s - 2n \mp 1 -\beta }
		- \frac{2s-2n \pm 1 -\beta}{2s} \left( \frac{1}{1+\gamma x} \right)
	\right]
	\mp \sigma_0^2 \frac{\textrm{d}}{ \textrm{d} x}
	\right\}
	\nonumber \\ 
	& 
	\times \left[ 
		\sqrt{{\frac{(2s-2n \mp 2-{\beta})(2s-n \mp 1-{\beta})^{\pm 1}}{
		 			 (2s-2n-{\beta})(2s-n-{\beta})^{\pm 1}}}}
	\frac{2s-2n \mp 1-{\beta}}{2s}
	\right] \psi_{n,\beta} (x).
\end{align}
The wavefunction for ground state
(\ref{eq:SI_ground_state}) 
is obtained from 
$\hat{L}_{\scriptscriptstyle -} (\beta)\psi_{0,\beta}(x) = 0,$
and the $n$th excited state is expressed by
\begin{equation}
\psi_{n,\beta} (x) = 
	\frac{1}{\sqrt{[n_\gamma ({\beta})]!}}
	[\hat{L}_{\scriptscriptstyle +}({\beta})]^n \psi_{0,\beta}(x).
\end{equation}

A dynamic group associated with the ladder operators
$\hat{L}_{\scriptscriptstyle -}$ and $\hat{L}_{\scriptscriptstyle +}$ 
can be built. 
For the case $\beta=1$, the action of the commutator 
$[\hat{L}_{\scriptscriptstyle -}, \hat{L}_{\scriptscriptstyle +}]$ 
on the eigenfunctions is
\begin{equation}
[\hat{L}_{\scriptscriptstyle -}, 
\hat{L}_{\scriptscriptstyle +}] \psi_n(x) = 
2l_{0}(n) \psi_n(x),
\end{equation}
with the eigenvalues defined by
$l_{0}(n) =\frac{1}{2} \left[ 1 - \gamma^2 \sigma_0^2 (n+1) \right],$
so that we can introduce the operator
$
\hat{L}_{0} =\frac{1}{2} [ 1 - \gamma^2 \sigma_0^2 (\hat{n}+1) ].
$
From the operators
$\hat{M}_{\scriptscriptstyle \pm} = \sqrt{2s} \hat{L}_{\scriptscriptstyle \pm}$
and
$\hat{M}_{0} = 2s \hat{L}_{0}$,
we get the following commutation relations
\begin{equation}
[\hat{M}_{\scriptscriptstyle +}, \hat{M}_{\scriptscriptstyle -}] =
2\hat{M}_{0}
\quad \textrm{and}\quad
[\hat{M}_{0}, \hat{M}_{\scriptscriptstyle \pm}] =
\pm \hat{M}_{\scriptscriptstyle \pm},
\end{equation}
which corresponds to a Lie algebra $SU(1,1)$ for
the deformed oscillator.

\section{\label{sec:coherent-states}
		 Coherent states}

\subsection{Coherent states in position representation}

Now we look for the quantum mechanical states for which 
the time evolution of the expected values of the  position 
and the momentum operators are similar to their respective classic analogues 
of the deformed oscillator.
Theses are the coherent states, 
whose classicity is expressed by 
the minimization of the uncertainty relationship 
(\ref{eq:uncertainty-relation}).
In order to achieve this goal we use the formalism introduced in 
Ref.~\onlinecite{Ruby-Senthilvelan-2010} for building coherent states 
of quantum systems with PDM.
The deformed annihilation and creation operators
(\ref{eq:a_gamma-and-a_gamma-dagger}) can be rewritten as
\begin{subequations}
\label{eq:creation-annihilation_m(x)}
\begin{align}
\hat{a}_\gamma & = 
		\sqrt{\frac{m_0 \omega_0}{2\hbar}}	
		\left[ \hat{\Phi}_\gamma
		+ \frac{\hbar}{m_0 \omega_0} 
			\sqrt[4]{\frac{m_0}{m(\hat{x})}} 
			\frac{\textrm{d}}{\textrm{d}\hat{x}} 
		   \sqrt[4]{\frac{m_0}{m(\hat{x})}} \,
		\right]
\\
\hat{a}_\gamma^{\dagger} &=
		\sqrt{\frac{m_0 \omega_0}{2\hbar}}	
		\left[ \hat{\Phi}_\gamma 
		- \frac{\hbar}{m_0 \omega_0} 
			\sqrt[4]{\frac{m_0}{m(\hat{x})}} 
			\frac{\textrm{d}}{\textrm{d}\hat{x}} 
		   \sqrt[4]{\frac{m_0}{m(\hat{x})}} \,  
		\right],
\end{align}
\end{subequations}
with 
$
\hat{\Phi}_\gamma 
= \hat{x} + \frac{\gamma \hbar}{2m_0 \omega_0}\hat{1}
$
the superpotential for the deformed oscillator.

In terms of the operators annihilation and creation,
the superpotential and the linear pseudo-momentum 
have the similar form of the standard oscillator
\begin{subequations}
\label{eq:Phi_gamma_Pi_gamma}
\begin{align}
\label{eq:Phi_gamma_a}
\hat{\Phi}_\gamma &= 
	\sqrt{\frac{\hbar}{2m_0 \omega_0}}	
	\left( \hat{a}_\gamma  + \hat{a}_\gamma^{\dagger} \right)
\\
\label{eq:Pi_gamma_a}
\hat{\Pi}_\gamma &= 
	i\sqrt{\frac{m_0 \hbar \omega_0}{2}}	
	\left( \hat{a}_\gamma^{\dagger} -\hat{a}_\gamma \right), 
\end{align}
\end{subequations}
and they satisfy the deformed commutation relation 
$
[\hat{\Phi}_\gamma, \hat{\Pi}_\gamma] 
= i\hbar \frac{\Phi'_\gamma (\hat{x})}{\sqrt{m(\hat{x})/m_0}}
= i\hbar (\hat{1} + \gamma \hat{x}).
$
Similar to the usual case, the superpotential is 
directly related to the wave function in the ground state.
In the case of the deformed oscillator, it turns out that
\begin{align}
\psi_0 (x)  &= \mathcal{N}_0 \sqrt[4]{\frac{m(x)}{m_0}}
			\exp \left[
			-\frac{1}{\sigma_0^2} \int^x
			\sqrt{\frac{m(y)}{m_0}}
			\Phi_\gamma (y) \textrm{d}y
			\right] \nonumber \\
		    &= 
			\frac{\mathcal{N}_0}{\sqrt{1+\gamma x}}
			\exp \left[
			-\frac{1}{\sigma_0^2} \int^x \Phi_\gamma (y) \textrm{d}_\gamma y
			\right]
\end{align}
recovers Eq.~(\ref{eq:ground-state})
for the superpotential (\ref{eq:Phi_gamma_a}).

Analogously as made by Glauber,\cite{Glauber-1963} 
the coherent states for a PDM particle are defined 
from the eigenvalues equation of
the annihilation operator\cite{Ruby-Senthilvelan-2010}
\begin{equation}
\hat{a}_\gamma |\alpha_\gamma \rangle 
= \alpha_\gamma| \alpha_\gamma \rangle, 
\end{equation}
where the wavefunction for the coherent states 
$\psi_{\textrm{cs}}(x) = \langle{x}|{\alpha_\gamma} \rangle$
in the position representation can be obtained 
by means of Eq.~(\ref{eq:a_gamma}), such that
\begin{equation}
\label{eq:CS-differential-equation}
\frac{1}{\sqrt{2}\sigma_0} \left( 
	x + \gamma \sigma_0^2 
	+\sigma_0^2 (1+\gamma{x})\frac{\textrm{d}}{\textrm{d}x}
	\right) \psi_{\textrm{cs}}(x) = \alpha_\gamma \psi_{\textrm{cs}}(x).
\end{equation}
Solving Eq.~(\ref{eq:CS-differential-equation}), we
arrived at a solution that is similar to the ground state 
(\ref{eq:ground-state}) 
\begin{equation}
\label{eq:psi_cs(x)}
\psi_{\textrm{cs}}(x) = 
		\mathcal{N}_{\textrm{cs}} \sqrt{2s}
		e^{-\frac{z(x)}{2}} [z(x)]^{\frac{\sqrt{2}\alpha_\gamma}{\gamma\sigma_0}
						+\frac{1}{\gamma^2\sigma_0^2}-1},
\end{equation}
with the normalization constant 
$
\mathcal{N}_{\textrm{cs}} =
 \sqrt{\gamma/\Gamma (\lambda_{\textrm{cs}})}
$
and 
$\lambda_{\textrm{cs}} 
	= \frac{2}{\gamma^2 \sigma_0^2} 
	  [1+\sqrt{2}\gamma \sigma_0 \textrm{Re}(\alpha_{\gamma})] -1.
$
The probability density for coherent states in position representation
is the Gamma distribution (\ref{eq:Gamma_distribuition}) 
with $\lambda$ replaced by $\lambda_{\textrm{cs}}$, i.e.,
$
\rho_{\textrm{cs}} (x) =
	\frac{2}{\gamma \sigma_0^2 \Gamma (\lambda_{\textrm{cs}})}
	e^{-z(x)} [z(x)]^{\lambda_{\textrm{cs}}-1}.
$

Alternatively, it is possible obtain the coherent states for PDM 
from the Perelomov approach,\cite{Ruby-Senthilvelan-2010} 
which is based on the definition of a displacement operator 
$\hat{\Xi}_\gamma$, 
such that
$|\alpha_\gamma \rangle = \hat{\Xi}_\gamma(\alpha_\gamma) |0\rangle,$
with
$\hat{\Xi}_\gamma (\alpha_\gamma) = e^{i\hat{\chi}_\gamma ( \alpha_\gamma )}$
and
\begin{equation}
\hat{\chi}_\gamma ( \alpha_\gamma ) = -i   \frac{\sqrt{2} \alpha_\gamma}{\sigma_0} 
	\int^{\hat{x}} \sqrt{\frac{m(\hat{y})}{m_0}} \textrm{d}\hat{y}
	= -i\sqrt{2} \alpha_\gamma \frac{\hat{x}_\gamma}{\sigma_0}.
\end{equation}
In fact,
\begin{align}
\psi_{\textrm{cs}} (x) 
	&= \langle x| \hat{\Xi}_\gamma (\alpha_\gamma) |0\rangle \nonumber \\
	&\propto   \exp \left[ \sqrt{2} \alpha_\gamma 
		\frac{\ln (1+ \gamma x)}{\gamma \sigma_0}
	  \right] \psi_0(x) \nonumber \\
	& \propto  e^{-\frac{1+\gamma x}{\gamma^2 \sigma_0^2}} 
	(1+\gamma x)^{\frac{\sqrt{2} \alpha_\gamma}{\gamma \sigma_0} 
				  + \frac{1}{\gamma^2 \sigma_0^2} - 1}
\end{align}
has the same form as Eq.~(\ref{eq:psi_cs(x)}).
Likewise the usual case, the displacement operator satisfies
$
\hat{\Xi}_\gamma^{\dagger} (\alpha_\gamma) 
\hat{a}_\gamma \hat{\Xi}_\gamma (\alpha_\gamma)
 = \hat{a}_\gamma + \alpha_\gamma
$
and
$
\hat{\Xi}_\gamma^{\dagger} (\alpha_\gamma) 
\hat{a}_\gamma^{\dagger} \hat{\Xi}_\gamma (\alpha_\gamma)
 = \hat{a}_\gamma^{\dagger} + \alpha_\gamma^{\ast},
$
with
$[\hat{\chi}(\alpha_\gamma), \hat{a}_\gamma] = i\alpha_\gamma$
and
$[\hat{\chi}(\alpha_\gamma), \hat{a}_\gamma^{\dagger}] = i\alpha_\gamma^{\ast}$.

\subsection{Minimum relation of uncertainty with the coherent states}

The expected values 
$\langle \hat{x} \rangle_{\textrm{cs}},$
$\langle \hat{x}^2 \rangle_{\textrm{cs}},$
$\langle \hat{\Pi}_\gamma \rangle_{\textrm{cs}},$
$\langle \hat{\Pi}_\gamma^2 \rangle_{\textrm{cs}}$
for coherent states are
\begin{subequations}
\label{eq:cs_expected_values}
\begin{align}
\label{eq:cs_expected_values_x}
& \langle \hat{x} \rangle_{\textrm{cs}} = 
	\frac{\sigma_0}{\sqrt{2}} 
	(\alpha_{\gamma}^{\ast} + \alpha_{\gamma})
	-\frac{\gamma \sigma_0^2}{2},  \\
\label{eq:cs_expected_values_x^2}
& \langle \hat{x}^2 \rangle_{\textrm{cs}} = 
	\frac{\sigma_0^2}{2} \left[ 
	1+(\alpha_{\gamma}^{\ast} + \alpha_{\gamma})^2 
	- \frac{\gamma \sigma_0}{\sqrt{2}} 
	(\alpha_\gamma^{\ast} + \alpha_\gamma) \right], \\
\label{eq:cs_expected_values_Pi}
& \langle \hat{\Pi}_\gamma \rangle_{\textrm{cs}} = 
	\frac{i\hbar}{\sqrt{2}\sigma_0} 
	(\alpha_{\gamma}^{\ast} - \alpha_{\gamma}), \\
\label{eq:cs_expected_values_Pi^2}
& \langle \hat{\Pi}_\gamma^2 \rangle_{\textrm{cs}} =
	\frac{\hbar^2}{2 \sigma_0^2} 
	\left[ 1-(\alpha_{\gamma}^{\ast} - \alpha_{\gamma})^2 
	+ \frac{\gamma \sigma_0}{\sqrt{2}} 
	(\alpha_\gamma^{\ast} + \alpha_\gamma)
	- \frac{\gamma^2 \sigma_0^2}{2} \right].
\end{align}
\end{subequations}
From Eqs.~(\ref{eq:cs_expected_values_x})--(\ref{eq:cs_expected_values_Pi^2}),
we obtain the uncertainty relations of 
the position and the linear pseudo-momentum
$
(\Delta x)_{\textrm{cs}}^2 = 
		\frac{\sigma_0^2}{2} 
		(1 + \gamma \langle \hat{x} \rangle_{\textrm{cs}})
$
and
$
(\Delta \Pi_{\gamma})_{\textrm{cs}}^2 = 
		\frac{\hbar^2}{2\sigma_0^2} 
		(1 + \gamma \langle \hat{x} \rangle_{\textrm{cs}}).
$
Therefore,
\begin{equation}
(\Delta x)_{\textrm{cs}}^2 (\Delta \Pi_\gamma)_{\textrm{cs}}^2
= \frac{\hbar^2}{4} (1 + \gamma \langle \hat{x} \rangle_{\textrm{cs}} )^2
= \frac{1}{4} |\langle [\hat{x}, \hat{\Pi}_{\gamma}] \rangle |_{\textrm{cs}}^2,
\end{equation}
i.e., the coherent states $|\alpha_\gamma \rangle$ minimize
the generalized uncertainty relation (\ref{eq:uncertainty-relation})
for the deformed oscillator.

\begin{figure*}[!bht]
\centering
\includegraphics[width=0.32\linewidth]{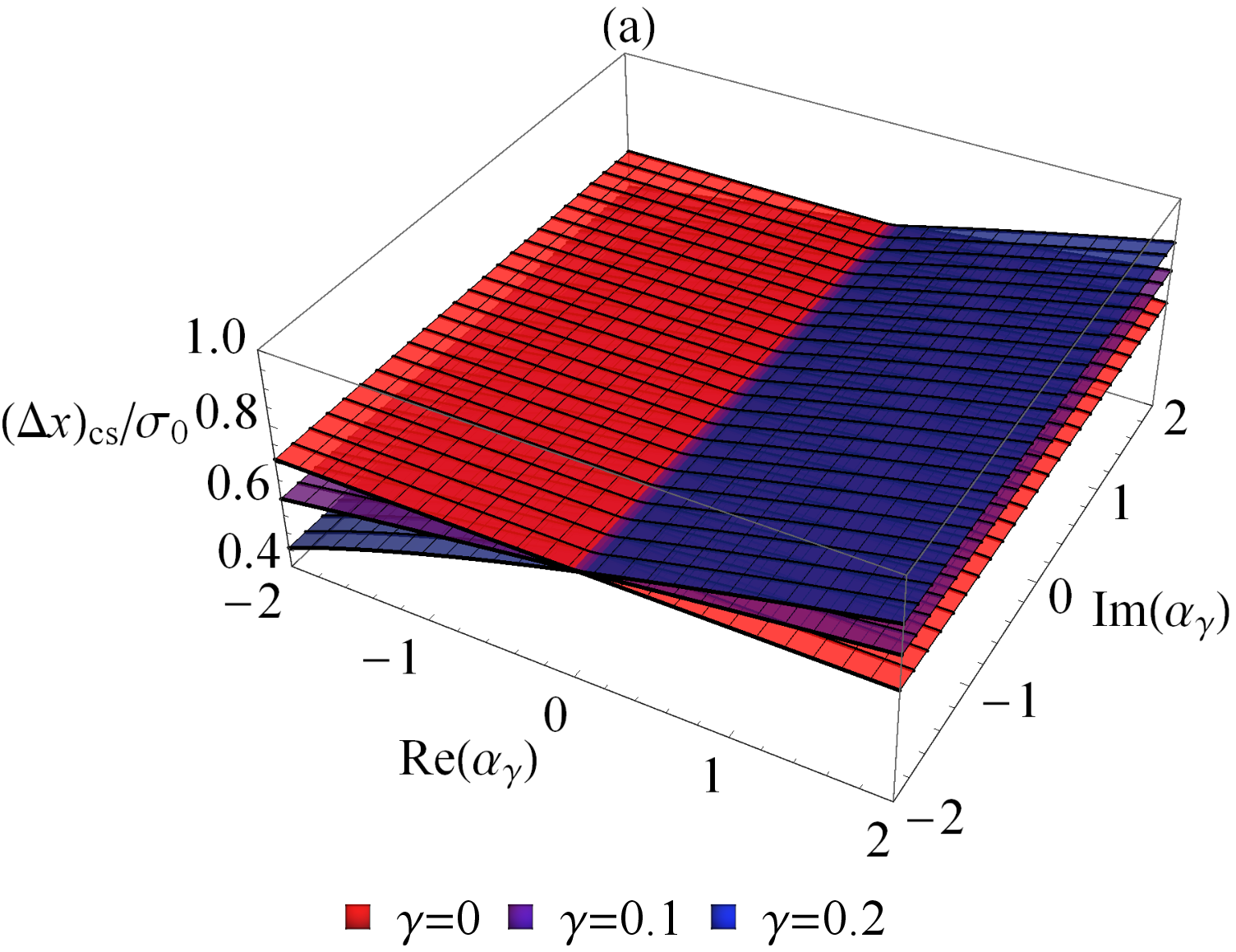}
\includegraphics[width=0.32\linewidth]{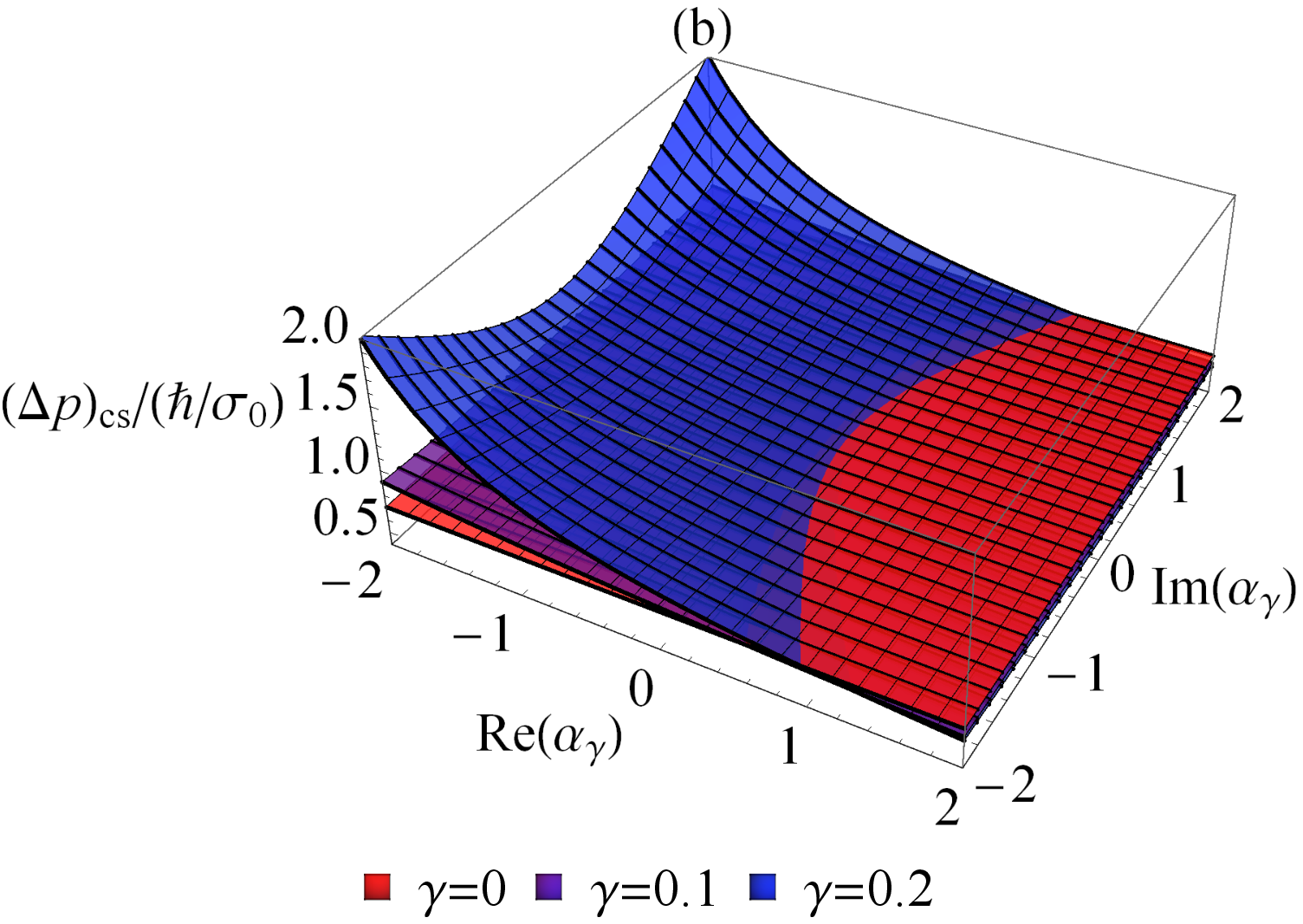}
\includegraphics[width=0.32\linewidth]{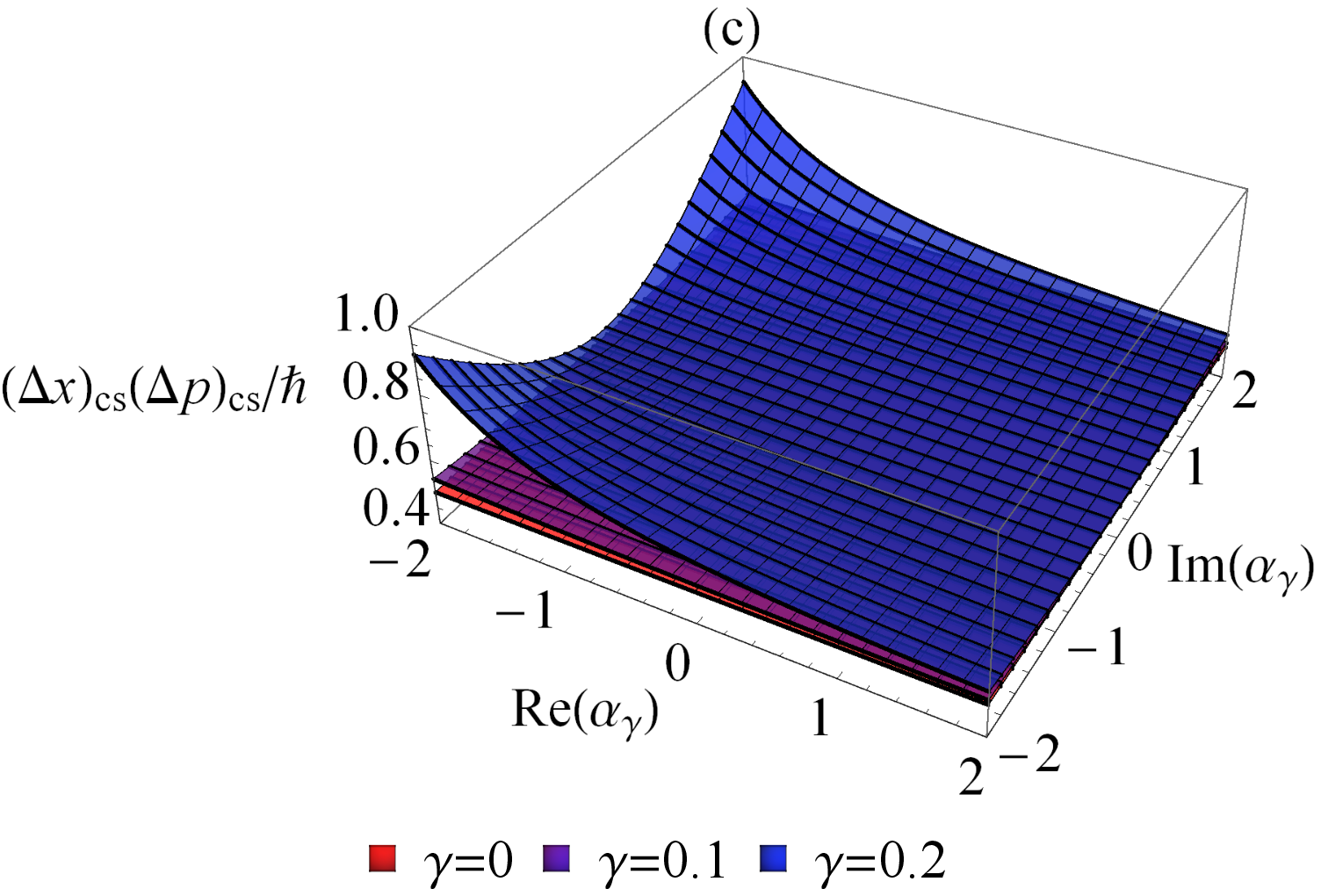}
\caption{\label{fig:2}
Uncertainties of 
(a) the position $(\Delta x)_{\textrm{cs}}$ and
(b) the linear momentum $(\Delta p)_{\textrm{cs}}$ 
along with the product 
(c) $(\Delta x)_{\textrm{cs}} (\Delta p)_{\textrm{cs}}$ 
for the coherent states of the deformed oscillator
in function of the complex number $\alpha_\gamma$, 
within the region 
$\{ |\textrm{Re}(\alpha_\gamma)|,|\textrm{Im}(\alpha_\gamma)|<2 \}$.
The surfaces are illustrated for three representative
values of the deformation parameter:
$\gamma \sigma_0 = 0$ (usual case -- red planes),
$0.1$ (violet surfaces) and
$0.2$ (blue surfaces).
}
\end{figure*}

It is straightforward to verify that the expectation values 
of the momentum linear satisfy
\begin{subequations}
\label{eq:cs_expected_values_p_and_p^2}
\begin{align}
\label{eq:cs_expected_values_p}
\langle \hat{p} \rangle_{\textrm{cs}}
	&= \frac{i\hbar}{{\sqrt{2}\sigma_0}}
	   \frac{(\alpha_\gamma^{\ast}-\alpha_\gamma )}{1
		+\frac{\gamma \sigma_0}{\sqrt{2}} (\alpha_\gamma^{\ast}
		+\alpha_\gamma) -\gamma^2 \sigma_0^2},
\\
\label{eq:cs_expected_values_p^2}
\langle \hat{p}^2 \rangle_{\textrm{cs}}
	&= \frac{\hbar^2}{2\sigma_0^2}
	   \frac{1}{1 
		+\frac{\gamma \sigma_0}{\sqrt{2}} (\alpha_\gamma^{\ast} +\alpha_\gamma) 
		-\frac{3}{2}\gamma^2 \sigma_0^2}
		\left[1 - \frac{(\alpha_\gamma^{\ast}-\alpha_\gamma )^2}{1 
		+\frac{\gamma \sigma_0}{\sqrt{2}} (\alpha_\gamma^{\ast} +\alpha_\gamma) 
		-\gamma^2 \sigma_0^2} \right].
\end{align}
\end{subequations}
In Fig.~\ref{fig:2},
from Eqs.~(\ref{eq:cs_expected_values_x}), (\ref{eq:cs_expected_values_x^2})
and (\ref{eq:cs_expected_values_p_and_p^2}) we plot the uncertainties
of the position $(\Delta x)_{\textrm{cs}}$ 
and of the linear momentum $(\Delta p)_{\textrm{cs}}$
along with the product $(\Delta x)_{\textrm{cs}}(\Delta p)_{\textrm{cs}}$
for the coherent states $|\alpha_\gamma \rangle$
that belong to the region 
$|\textrm{Re}(\alpha_\gamma)|,|\textrm{Im}(\alpha_\gamma)|<2$.
These three quantifiers result symmetric around 
$\textrm{Im}(\alpha_\gamma)=0$.
The plane of the uncertainty relation 
(corresponding to the standard case $\gamma \sigma_0=0$)
becomes a curved surface as the deformation parameter
$\gamma \sigma_0$ grows, thus allowing to distinguish 
the coherent states within the range analyzed.

\subsection{Quasi-classical states}

Applying the time evolution operator
$\hat{\mathcal{U}}(t,t_0) = e^{-i\hat{H}(t-t_0)/\hbar}$ on equation
$|\alpha_\gamma (t_0) \rangle = 
\hat{\Xi}_\gamma(\alpha_\gamma (t_0)) |0\rangle,$
and considering the Heisenberg picture for
displacement operator 
--- $\hat{\Xi}_\gamma (t) = 
\hat{\mathcal{U}}(t,t_0) 
	\hat{\Xi}_\gamma (t_0) 
\hat{\mathcal{U}}^{\dagger}(t,t_0)$ --- 
we get the coherent states at the instant $t$ is
$|\alpha_\gamma (t) \rangle = e^{-iE_0(\gamma)(t-t_0)/\hbar} 
\hat{\Xi}_\gamma(\alpha_\gamma (t)) |0\rangle.$
So, to go from $|\alpha_\gamma (t_0) \rangle$ to
$|\alpha_\gamma (t) \rangle$,
it is only necessary to replace $\alpha_\gamma(t_0)$ 
by $\alpha_\gamma(t)$,
and multiply the result by the phase factor
$ e^{-iE_0(\gamma)(t-t_0)/\hbar}$.
The corresponding probability density is given by
\begin{equation}
\label{eq:rho_cs(x,t)}
\rho_{\textrm{cs}} (x, t) 
		= \frac{2}{\gamma \sigma^2 \Gamma (\lambda_{\textrm{cs}} (t))}
		   e^{-z(x)} [z(x)]^{\lambda_{\textrm{cs}}(t)  -1}.
\end{equation}
with shape parameter
$
\lambda_{\textrm{cs}} (t) 
	= \frac{2}{\gamma^2 \sigma_0^2} 
	  (1+ \gamma A \cos \Theta_\gamma (t)) -1
$
and $\alpha_\gamma (t) = |\alpha_\gamma| e^{-i \Theta_\gamma (t)}$,
in which $\Theta_\gamma (t)$ is a deformed phase for coherent states.

\begin{figure}[!hbt]
\centering
\includegraphics[width=0.38\linewidth]{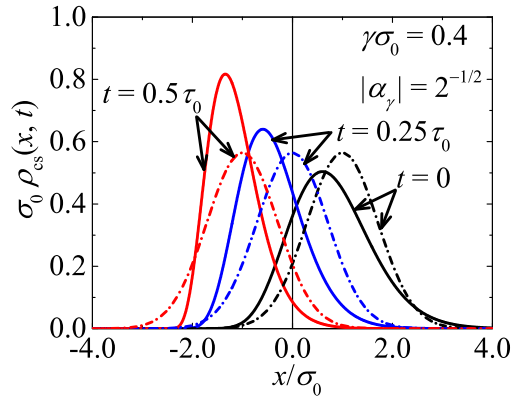}
\caption{\label{fig:3}
Motion of the probability density
associated to the coherent states 
$| \alpha_\gamma \rangle$ of the deformed oscillator with 
the mass function $m(x) = m_0/(1+\gamma x)^2$ at the times 
$t=0$, $\tau_0/4$ and $\tau_0/2$ ($\tau_0 = 2\pi/\omega_0$)
for $\gamma \sigma_0 = 0.4$ and $|\alpha_\gamma| = 1/\sqrt{2}$.
The usual case ($\gamma = 0$, dashed line) is also illustrated 
for comparison.
}
\end{figure}

The time evolution of the operator $\hat{a}_\gamma$ is
\begin{equation}
	\frac{\textrm{d}}{\textrm{d}t}\hat{a}_\gamma
	= \frac{1}{i\hbar} [\hat{a}_\gamma, \hat{H}]
	= -i\omega_0 (1+\gamma \hat{x})\hat{a}_\gamma.
\end{equation}
Let 
$\hat{a}_\gamma (t) = 
\hat{a}_\gamma (t_0) e^{-i\Theta_\gamma (t)}$,
then we arrive at
\begin{equation}
\label{eq:dot_Theta}
\dot{\Theta}_\gamma (t)\hat{a}_\gamma (t) = 
\omega_0 (\hat{1}+\gamma \hat{x}(t) )\hat{a}_\gamma (t).
\end{equation}
For coherent states, Eq.~(\ref{eq:dot_Theta}) can be rewritten as 
$
\dot{\Theta}_\gamma (t)
 = \omega_0 \left( 
		1 + \gamma A \cos \Theta_\gamma 
		- \frac{\gamma^2 \sigma_0^2}{2} 
	\right).
$	
Analogously to the classical formalism, 
by integrating we obtain
\begin{equation}
\label{eq:deformed-phase_CS}
\Theta_\gamma (t) = 
	2\textrm{tan}^{-1} \left\{ 
	\sqrt {\frac{1 + \gamma A_{\textrm{cs},\gamma}}{1-\gamma A_{\textrm{cs},\gamma}}}
	\textrm{tan} \left[ 
	\frac{1}{2} \Omega_{{\textrm{cs}},\gamma} (t-t_0)
	\right] \right\}
\end{equation}
with 
$\Theta_\gamma(t_0)=0$
and the parameters modified 
$A_{\textrm{cs},\gamma} = 
A\left(1-\frac{\gamma^2 \sigma_0^2}{2}\right)^{-1}$
and
$\Omega_{\textrm{cs},\gamma} = \omega_0 
\sqrt{\left(1-\frac{\gamma^2 \sigma_0^2}{2}\right)^2 - \gamma^2 A^2}.$
Figure~\ref{fig:3} shows a comparison between 
the motion of the probability densities 
$\rho_{\textrm{cs}} (x, t)$ [Eq.~(\ref{eq:rho_cs(x,t)})]
for cases $\gamma \sigma_0 = 0$ (standard oscillator) and $0.4$.

\begin{figure}[!htb]
\centering
\includegraphics[width=0.38\linewidth]{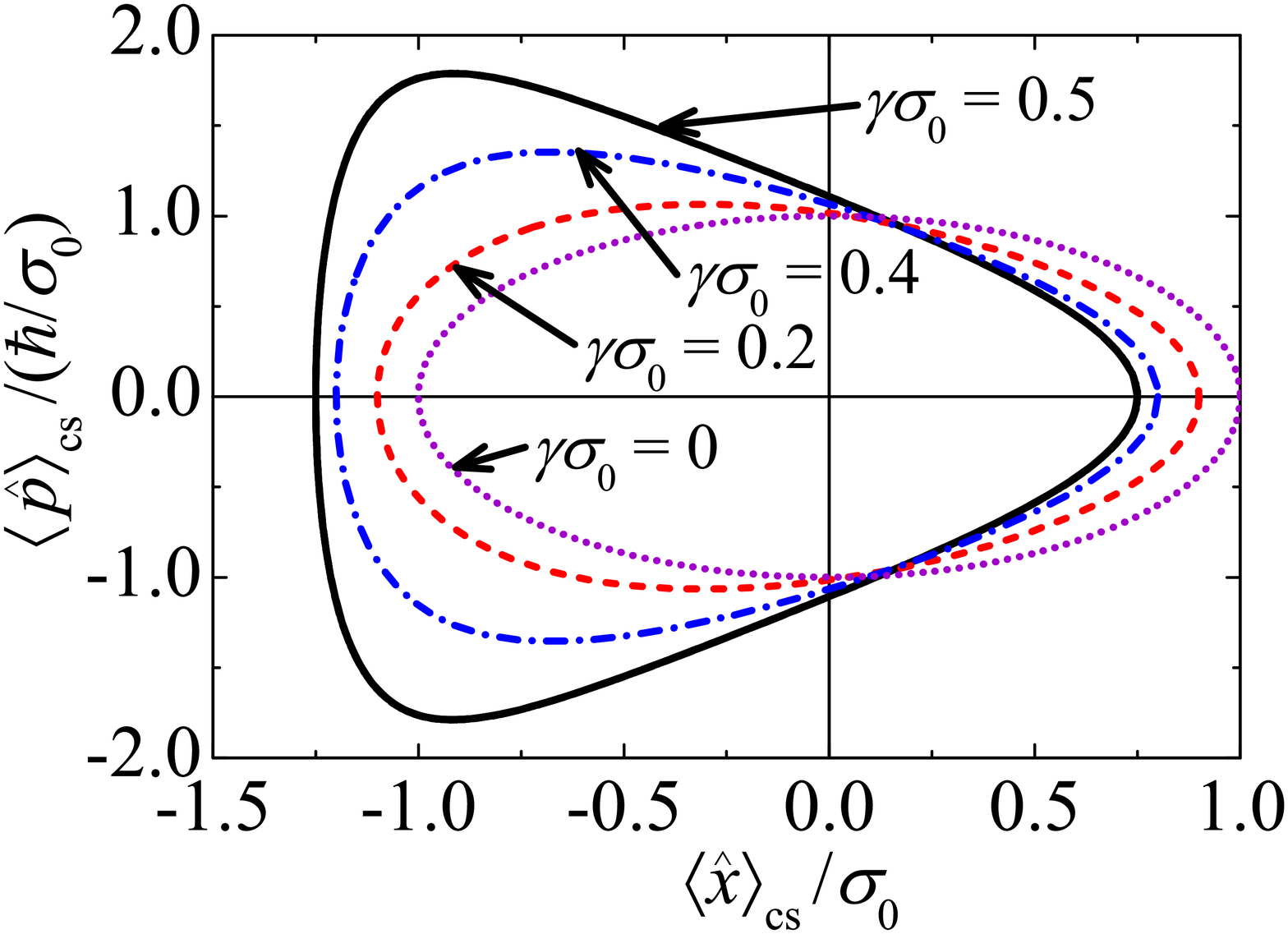}
\caption{\label{fig:4}
Phase space of the deformed oscillator associated 
to the coherent states $|\alpha_\gamma \rangle$
for $|\alpha_\gamma| = 1/\sqrt{2}$ and 
deformation parameter 
$\gamma \sigma_0=0$ (dotted purple -- usual case), 
$0.2$ (dashed red), 
$0.4$ (dashed-dotted blue) 
and $0.5$ (solid black).
}
\end{figure}

Since
$ \textrm{Re} [\alpha_\gamma (t)] =  
|\alpha_\gamma| \cos \Theta_\gamma (t)$
and
$ \textrm{Im} [\alpha_\gamma (t)] =  
|\alpha_\gamma| \sin \Theta_\gamma (t),$
using Eqs.~(\ref{eq:cs_expected_values_x}) 
and (\ref{eq:cs_expected_values_Pi}) 
we can express
\begin{subequations}
\begin{align}
\label{eq:x_cs}
\langle \hat{x}  (t) \rangle_{\textrm{cs}}
&= A\cos{\Theta_\gamma} (t)- \frac{\gamma\sigma_0^2}{2},
\\
\label{eq:Pi_gamma_cs}
\langle \hat{\Pi}_\gamma (t) \rangle_{\textrm{cs}} 
&= - m_0 \omega_0 A\sin{\Theta_\gamma} (t).
\end{align}
\end{subequations}
As expected, the mean values of the position and 
the linear momentum 
(Eqs.~(\ref{eq:x_cs})--(\ref{eq:Pi_gamma_cs}))
evolve in the same way as their classic analogues.
The position (\ref{eq:x_cs}) oscillates around 
an equilibrium position 
$x_{\textrm{eq},\gamma} = - \frac{\gamma \sigma_0^2}{2}$
due to the uniform electric field
produced by the deformation of the space, 
whose force is $F_\gamma
= - \textrm{d} \mathcal{V}_\gamma(x)/\textrm{d}x
= - \frac{1}{2}\hbar \omega_0 \gamma$.
From Eq.~(\ref{eq:cs_expected_values_p}),
the linear momentum is
\begin{equation}
\label{eq:p_gamma_cs}
\langle \hat{p} (t) \rangle_{\textrm{cs}} 
= - \frac{m_0 \omega_0 A\sin{\Theta_\gamma} (t)}{
	1+\gamma A \cos \Theta_\gamma (t) - \gamma^2 \sigma_0^2}.
\end{equation}
Figure~\ref{fig:4} illustrates the phase space from 
the expected values (\ref{eq:x_cs}) and (\ref{eq:p_gamma_cs})
for different values of $\gamma \sigma_0$.
The paths are similar to corresponding classical analogues 
investigated previously in Ref.~\onlinecite{Costa-Borges-2018}. 
When $\gamma$ increases the center of 
the trajectory is shifted to the left.

In according to Ehrenfest's theorem,
the expected values of the superpotential and pseudo-momentum 
satisfy equations of motion identical to the classical analogues 
(\ref{eq:classical_equation_of_motion}), 
\begin{equation}
\label{eq:quantum_equation_of_motion}
\left\{
\begin{array}{rcl}
\displaystyle
\frac{1}{\sigma_0} \displaystyle 
\mathcal{D}_\gamma  
\langle \hat{\Phi}_\gamma (t) \rangle_{\textrm{cs}}   &=& 
\displaystyle
\frac{\sigma_0}{\hbar} \langle \hat{\Pi}_\gamma  (t) \rangle_{\textrm{cs}} \\
\displaystyle
\frac{\sigma_0}{\hbar} \displaystyle 
\mathcal{D}_\gamma  
\langle \hat{\Pi}_\gamma  (t) \rangle_{\textrm{cs}} 
&=& 
\displaystyle
- \frac{1}{\sigma_0}\langle \hat{\Phi}_\gamma  (t) \rangle_{\textrm{cs}}
\end{array}
\right.
\end{equation}
with 
$\mathcal{D}_\gamma (\,\cdot\,) = 
\frac{1}{1 + \gamma \langle \hat{x} (t) \rangle_{\textrm{cs}}} 
\frac{\textrm{d}}{\textrm{d}t} (\,\cdot\,).
$
 
The expected values of the Hamiltonian operator
for coherent states are
$	
\langle \hat{H} \rangle_{\textrm{cs}} 
		= \mathcal{H} + E_0({\gamma}).
$
In this way, we recover
$
|{\alpha}_{\gamma}|^2 =
\frac{\langle \hat{H} \rangle_{\textrm{cs}}-E_0(\gamma)}{\hbar \omega_0}
= \frac{A^2}{2\sigma_0^2}.
$
If $|\alpha_\gamma|$ is very large, then $A \gg \sigma_0$, 
and the quantum oscillator phase of the coherent states
behave like the classical oscillator phase 
$\Theta_{\gamma} (t) \approx \theta_\gamma(t)$,
as well as
$\langle \hat{\Phi}_\gamma (t) \rangle_{\textrm{cs}} \rightarrow x (t)$
and
$\langle \hat{\Pi}_\gamma (t) \rangle_{\textrm{cs}} \rightarrow \Pi_\gamma (t)$.
As predicted, the dynamics of the coherent states 
for deformed oscillator are similar to the 
classical equations described 
Section \ref{sec:classical-and-quantum-deformed-formalism}. 

\begin{figure}[!htb]
\centering
\includegraphics[width=0.38\linewidth]{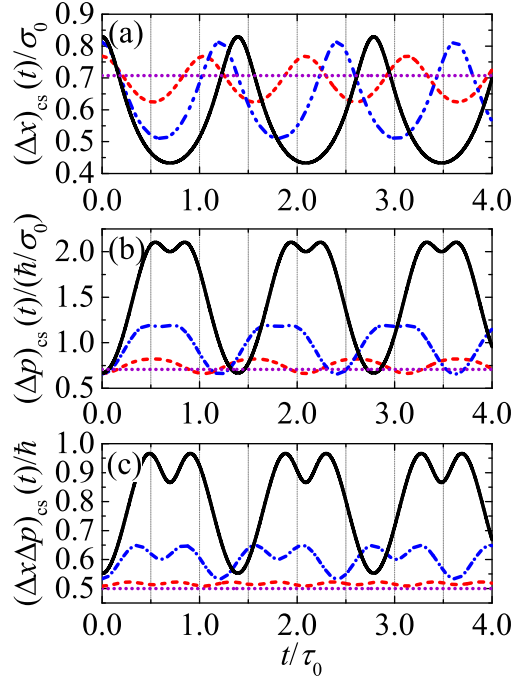}
\caption{\label{fig:5}
Time evolution of the uncertainties of 
(a) the position $(\Delta x)_{\textrm{cs}}(t)$ and 
(b) the linear momentum $(\Delta p)_{\textrm{cs}}(t)$, and 
(c) of the product $(\Delta x)_{\textrm{cs}}(t)(\Delta p)_{\textrm{cs}}(t)$
for the coherent states of the deformed oscillator with PDM,
being $|\alpha_\gamma| = 1/\sqrt{2}$ and deformation parameters
$\gamma \sigma_0=0$ (dotted purple -- usual case), 
$0.2$ (dashed red), 
$0.4$ (dashed-dotted blue) and
$0.5$ (solid black).
}
\end{figure}

From  $\alpha_\gamma (t) = |\alpha_\gamma| e^{-i \Theta_\gamma (t)}$,
Eqs.~(\ref{eq:cs_expected_values_x})--(\ref{eq:cs_expected_values_x^2}),
(\ref{eq:cs_expected_values_p})--(\ref{eq:cs_expected_values_p^2})
and deformed phase (\ref{eq:deformed-phase_CS}),
we plot in Fig.~\ref{fig:5}
the time evolution of the uncertainty relation 
[$(\Delta x)_{\textrm{cs}}(t) (\Delta p)_{\textrm{cs}}(t)$],
along the uncertainties 
$
(\Delta x)_{\textrm{cs}}(t) = 
\sqrt{\langle \hat{x}^2  (t) \rangle_{\textrm{cs}}
      -\langle \hat{x}  (t) \rangle_{\textrm{cs}}^2}
$
and
$
(\Delta p)_{\textrm{cs}}(t) = 
\sqrt{\langle \hat{p}^2  (t) \rangle_{\textrm{cs}}
      -\langle \hat{p}  (t) \rangle_{\textrm{cs}}^2}
$
for different values of $\gamma \sigma_0$.
When $\gamma \rightarrow 0$ the oscillatory behavior on the uncertainties 
of the position and linear momentum disappears, so that they becomes 
$(\Delta x)_{\textrm{cs}} = \frac{\sigma_0}{\sqrt{2}}$ 
and
$(\Delta p)_{\textrm{cs}} = \frac{\hbar}{\sqrt{2}\sigma_0}.$

\section{\label{sec:final-remarks}
	     Final remarks}

In this work, quantum and classical harmonic oscillators suitably 
deformed for accounting a particle with a position-dependent mass  
have been studied from the supersymmetric framework. 
We summarize our contributions as follows.
\begin{enumerate}[(i)]
\item
The classical and quantum Hamiltonians of 
the deformed harmonic oscillator are factorized in terms 
of their corresponding deformed ladder operators, 
the latter obtained by simply replacing the momentum 
by its deformed version.
\item
The classical and quantum ladder operators preserve the Poisson 
and Lie brackets structure as well as the Jacobi identity.
\item
The probability density of the ground state behaves like a Gamma distribution, 
as a result of the deformation, and it recovers the Gaussian package 
when the deformation tends to zero. 
\item
The position-dependent mass causes an additional term 
in the supersymmetric Hamiltonian, which is equivalent 
to introduce an uniform electric field in the $x$-direction. 
In turn, this feature could be used to represent uniform field 
interactions in terms of a PDM particle.
\item
The deformed formalism preserves the structure of the
shape invariant partner Hamiltonians in terms of the deformed 
ladder operators and the pseudo-momentum operator.
\item
The coherent states of the deformed harmonic oscillator
have the same structure than in the standard non-deformed case, 
by satisfying a minimum deformed uncertainty relation.
\item 
When the deformation is present the plane of the uncertainty relation 
transforms into a curved surface, thus serving as a distinguishability measure
for the coherent states. The distinguishability
becomes more pronounced as the the deformation parameter increases (Fig. 2).
\end{enumerate}

The deformed oscillator carries pieces of information about 
the corrections of the generalized uncertainty principle (GUP). 
This is reflected by an increasing of the peaks 
in the temporal evolution of $\Delta x\Delta p$ 
along with an anharmonic oscillatory behavior, 
as the deformation parameter grows (Fig. 4). 
We also see that for some times the minimum of 
$\Delta x\Delta p$ is very close to the one corresponding to 
the standard case ($\gamma \sigma_0=0$), 
being almost independent of the deformation parameter for 
the range of values analyzed. 
This can be physically interpreted as if the GUP corrections 
are periodically cancelled for a particle with 
PDM being in a coherent state.

Overall, the deformation addressed in this paper, 
inspired by the formalism of the non-additive quantum mechanics
\cite{CostaFilho-Almeida-Farias-AndradeJr-2011,
Mazharimousavi-2012,
Costa-Borges-2014,
Barbagiovanni-Costafilho-2013,
Barbagiovanni-2014,
Costa-Gomez-Santos-2020,
CostaFilho-Alencar-Skagerstam-AndradeJr-2013,
Costa-Borges-2018,
Costa-Gomez-Borges-2020,
Tchoffo-Vubangsi-Fai-2014,
Merad-etal_2019,
Arda-Server,
Aguiar-Cunha-daCosta-CostaFilho-2020,
CostaFilho-Oliveira-Aguiar-DaCosta-2021}
contains a structural richness that allows modeling, 
ranging from effective position-dependent masses, uniform field interactions, 
deformed coherent states, to corrections of the GUP. 
In this sense, the use of other deformations
could be helpful for modelling different scenarios.

The SUSY formalism applied to the deformed oscillator 
also opens new perspectives for the construction 
of other types of coherent states, 
such as the Barut-Girardello or
Gazeau-Klauder formalisms.\cite{Gazeau-2009}

\section*{Acknowledgments}
I. S. G. acknowledges support from
Coordena\c{c}\~ao de Aperfei\c{c}oamento de Pessoal de N\'ivel Superior (CAPES)
and Conselho Nacional de Desenvolvimento Cient\'ifico e Tecnol\'ogico
(CNPq -- Postdoctoral Fellowship 159799/2018-0),
Brazilian agencies.

\section*{Data availability}

Data sharing is not applicable to this article 
as no new data were created or analyzed in this study.

\renewcommand{\theequation}{A\arabic{equation}}
\setcounter{equation}{0}
\section*{Appendix}

In the following, we demonstrate the expressions
(\ref{eq:a_gamma-and-adjunct-eigenfunctions}).
Making the change of variable 
$x\rightarrow z = 2(1+\gamma x)/\gamma^2 \sigma_0^2,$
so that the annihilation and creation operators 
on wavefunctions can be written as
\begin{subequations}
\begin{align}
\label{eq:a_gamma_z}
&\hat{a}_\gamma \psi_n^{\scriptscriptstyle (+)}
	= \frac{\gamma \sigma_0}{\sqrt{2}}	
		\left( 
			1-\frac{1}{\gamma^2 \sigma_0^2} + \frac{z}{2}
			+z\frac{\textrm{d}}{\textrm{d}z}
		\right) \psi_n^{\scriptscriptstyle (+)},
		\\
\label{eq:a_gamma_dagger_z}
& \hat{a}_\gamma^{\dagger} \psi_n^{\scriptscriptstyle (-)}
	= \frac{\gamma \sigma_0}{\sqrt{2}}	
		\left( 
			-\frac{1}{\gamma^2 \sigma_0^2} + \frac{z}{2}
			-z\frac{\textrm{d}}{\textrm{d}z}
		\right) \psi_n^{\scriptscriptstyle (-)}.
\end{align}
\end{subequations}

Substituting in Eq.\;(\ref{eq:a_gamma_z}) 
the eigenfunctions of the deformed oscillator
in terms of the variable $z$,
$\psi_n^{\scriptscriptstyle (+)}(x(z)) =
(-1)^{n}\sqrt{2s} {\mathcal{N}}_n e^{-\frac{z}{2}} 
z^\frac{{\nu}_n-1}{2} L_n^{({\nu}_n)} (z),$ 
we get
\begin{equation}
\hat{a}_\gamma \psi_n^{\scriptscriptstyle (+)}
	= (-1)^{n}\mathcal{N}_n e^{-\frac{z}{2}} 
		z^{\frac{{\nu}_n -1}{2}}
			\left( 
			-n L_{n}^{({\nu}_n)}(z) 
			+ z \frac{\textrm{d}L_{n}^{({\nu}_n)}}{\textrm{d}z}
		\right).
\end{equation}
From the property of the Laguerre polynomials 
\begin{align}
\label{eq:Laguerre_property}
z \frac{\textrm{d}L_{n}^{(\nu)}(z)}{\textrm{d}z}
	&= (n+1) L_{n+1}^{(\nu)}(z) -(n+{\nu}+1-z)L_{n}^{(\nu)}(z),
	\nonumber \\
	&= n L_{n}^{(\nu)}(z) -(n+{\nu})L_{n-1}^{(\nu)}(z),
\end{align}
and using recurrence relationships
$
\mathcal{N}_n = \sqrt{\frac{n}{n+\nu_n}} \widetilde{\mathcal{N}}_{n-1}
$
and
$\nu_{n}=\widetilde{\nu}_{n-1},$
we arrive at
\begin{equation}
\hat{a}_\gamma \psi_n^{\scriptscriptstyle (+)} 
	= (-1)^{n-1}\sqrt{n(n+{\nu}_n)}
	   \widetilde{\mathcal{N}}_{n-1} 
	   e^{-\frac{z}{2}}z^{\frac{\widetilde{\nu}_{n-1}-1}{2}}
       L_{n-1}^{(\widetilde{\nu}_{n-1})}(z).
\end{equation}
Since 
$\sqrt{n(n+\nu_n)}=\sqrt{ 2s E_n^{\scriptscriptstyle (+)}/\hbar \omega_0},$
we obtain Eq.~(\ref{eq:a_gamma-eigenfunctions}).

Similarly, substituting 
in Eq.~(\ref{eq:a_gamma_dagger_z}) 
the eigenfunctions
$
\psi_{n}^{\scriptscriptstyle (-)}(x(z))= 
	(-1)^{n}\sqrt{2s} \widetilde{\mathcal{N}}_n e^{-\frac{z}{2}} 
	z^{\frac{\widetilde{\nu}_{n}-1}{2}} 
	L_n^{(\widetilde{\nu}_n)} (z),
$
leads us at
\begin{equation}
\hat{a}_\gamma^{\dagger} \psi_n^{\scriptscriptstyle (-)}
	= (-1)^{n+1}\widetilde{\mathcal{N}}_n e^{-\frac{z}{2}} 
		z^{\frac{\widetilde{\nu}_n -1}{2}}  
		\left[
			(n + \widetilde{\nu}_{n} + 1 - z) L_{n}^{(\widetilde{\nu}_{n})}(z) 
			+ z \frac{\textrm{d}L_{n}^{(\widetilde{\nu}_{n})}}{\textrm{d}z}
		\right].
\end{equation}
From Eq.~(\ref{eq:Laguerre_property}),
$
\widetilde{\mathcal{N}}_n = \sqrt{\frac{n+\nu_n-1}{n+1}}\mathcal{N}_{n+1} 
$
and
$\widetilde{\nu}_n = \nu_{n+1},$ 
we obtain
\begin{equation}
\hat{a}_\gamma^{\dagger} \psi_n^{\scriptscriptstyle (-)} 
	=  (-1)^{n+1}  \sqrt{(n+1)(n+{\nu}_n-1)}
	   \mathcal{N}_{n+1} 
       e^{-\frac{z}{2}}z^{\frac{{\nu}_{n+1}-1}{2}}
	   L_{n+1}^{(\nu_{n+1})}(z).
\end{equation}
Since 
$\sqrt{(n+1)(n+{\nu}_n-1)}=\sqrt{ 2s E_n^{\scriptscriptstyle (-)}/\hbar \omega_0},$
we prove Eq.~(\ref{eq:a_gamma_dagger_z}).

\section*{References}


\end{document}